\documentclass[prl,aps,twocolumn,nofootinbib,preprintnumbers,superscriptaddress,preprintnumbers,balancelastpage,longbibliography]{revtex4-2}
\usepackage[utf8]{inputenc}
\usepackage[colorlinks=true,citecolor=blue,linkcolor=blue]{hyperref}
\usepackage[normalem]{ulem}
\usepackage{amsmath,amssymb, mathrsfs,xfrac}
\usepackage{epsfig}
\usepackage{graphicx}               
\usepackage{color}
\usepackage{slashed}
\usepackage{multirow}
\usepackage{placeins}
\usepackage{graphicx}
\usepackage[dvipsnames]{xcolor}
\usepackage{epstopdf}
\usepackage{soul}
\usepackage{tikz}
\usepackage[capitalise, english]{cleveref}
\usepackage{siunitx}
\usepackage{xspace}
\usepackage{booktabs}
\usetikzlibrary{trees}
\usepackage{graphicx}
\usepackage[export]{adjustbox}
\usetikzlibrary{decorations.pathmorphing}
\usetikzlibrary{decorations.markings}
\usepackage{aas_macros}
\usepackage{lettrine}
\usepackage{subfigure}
\usepackage{graphicx}
\usepackage{float}
\usepackage{amssymb}
\usepackage{url}

\usepackage{placeins}
\usepackage{textgreek}
\usepackage{booktabs, tabularx, array}
\newcolumntype{Y}{>{\raggedright\arraybackslash}X}

\LettrineTextFont{\itshape}
\usepackage{comment}

\newcommand\myshade{80}

\colorlet{mylinkcolor}{ForestGreen}

\colorlet{mycitecolor}{Aquamarine}

\colorlet{myurlcolor}{violet}

\hypersetup{
  linkcolor  = mylinkcolor!\myshade!black,
  citecolor  = mycitecolor!\myshade!black,
  urlcolor   = myurlcolor!\myshade!black,
  colorlinks = true
}

\definecolor{jblue}{RGB}{20,50,100}
\definecolor{npurple}{RGB} {153, 51, 204}
\definecolor{wred}{RGB}{217,0,56}
\definecolor{white}{RGB}{255,255,255}

\definecolor{korange}{RGB}{235, 80,  43}
\definecolor{korange2}{RGB}{245, 100,  63}
\definecolor{kyelloworange}{RGB}{255, 210,  110}
\definecolor{kyelloworange2}{RGB}{240, 170,  90}
\definecolor{kred}{RGB}{204,  102, 153}
\definecolor{kpurple}{RGB}{153,  61, 190}
\definecolor{kpurplelight}{RGB}{213,  161, 230}

 \definecolor{tobycolour}{rgb}{.5,.0,.5}

\DeclareSIUnit\year{yr}
\DeclareSIUnit\pc{pc}
\DeclareSIUnit\ergs{ergs}
\DeclareSIUnit\msun{\ensuremath{M_\odot}}
\allowdisplaybreaks

\makeatletter
\providecommand*{\diff}%
  {\command{\lmultau}{\ensuremath{L_\mu-L_\tau}\xspace}
\new@ifnextchar^{\DIfF}{\DIfF^{}}}
\def\DIfF^#1{%
  \mathop{\mathrm{\mathstrut d}}%
    \nolimits^{#1}\gobblespace}
\def\gobblespace{%
  \futurelet\diffarg\opspace}
\def\opspace{%
  \let\DiffSpace\!%
  \ifx\diffarg(%
    \let\DiffSpace\relax
  \else
    \ifx\diffarg[%
      \let\DiffSpace\relax
    \else
        \ifx\diffarg\{%
        \let\DiffSpace\relax
      \fi\fi\fi\DiffSpace}

\usepackage{tikz,xcolor,hyperref}

\definecolor{lime}{HTML}{A6CE39}
\DeclareRobustCommand{\orcidicon}{\hspace{-1mm}
	\begin{tikzpicture}
	\draw[lime, fill=lime] (0,0) 
	circle [radius=0.16] 
	node[white] {{\fontfamily{qag}\selectfont \tiny \,ID}};
	\draw[white, fill=white] (-0.0525,0.095) 
	circle [radius=0.007];
	\end{tikzpicture}
	\hspace{-3mm}
}

\foreach \x in {A, ..., Z}{\expandafter\xdef\csname orcid\x\endcsname{\noexpand\href{https://orcid.org/\csname orcidauthor\x\endcsname}
			{\noexpand\orcidicon}}
}

\usepackage{physics}


\newcommand{\mytitle}{Latest $\nu$s from the Earliest Galaxies: Probing neutrino self-interactions with JWST}

\begin{document}

	\title{\mytitle}

\author{Ranjini Mondol\orcidA{}}
\email{ranjini12m@gmail.com, ranjini.mondol@saha.ac.in}
\affiliation{Theoretical Physics Division, Saha Institute of Nuclear Physics, 1/AF, Bidhannagar, Kolkata 700064, India}
\affiliation{Centre for High Energy Physics, Indian Institute of Science, C.\,V.\,Raman Avenue, Bengaluru 560012, India}

\author{Priyank Parashari\orcidC{}}
\email{ppriyank@usc.edu}
\affiliation{Department of Physics \& Astronomy, University of Southern California, Los Angeles, CA, 90089, USA}
\affiliation{Centre for High Energy Physics, Indian Institute of Science, C.\,V.\,Raman Avenue, Bengaluru 560012, India}

\author{Anirban Das\orcidB{}}
\email{anirbandas.21@protonmail.com}
\affiliation{Theoretical Physics Division, Saha Institute of Nuclear Physics, 1/AF, Bidhannagar, Kolkata 700064, India}
\affiliation{Homi Bhabha National Institute, Training School Complex, Anushaktinagar, Mumbai 400094, India}

\author{Vera Gluscevic\orcidE{}}
\email{vera.gluscevic@usc.edu}
\affiliation{Department of Physics \& Astronomy, University of Southern California, Los Angeles, CA, 90089, USA}

\author{Ranjan Laha\orcidD{}}
\email{ranjanlaha@iisc.ac.in}
\affiliation{Centre for High Energy Physics, Indian Institute of Science, C.\,V.\,Raman Avenue, Bengaluru 560012, India}

	\date{\today}

\begin{abstract}
Our Universe serves as a complementary laboratory for particle physics interactions. One intriguing possibility is that neutrinos may have beyond-the-Standard-Model (BSM) self-interactions, which can delay their free streaming and have an observable effect across different redshifts. Such interactions can induce scale-dependent changes in the matter power spectrum. The James Webb Space Telescope (JWST) has opened up a new window onto the early Universe by discovering a large population of high-redshift galaxies up to $z \sim 15$ and various probable galaxy candidates at even higher redshifts, between $z \sim 15-30$. Such high-redshift observations can probe the matter power spectrum on small scales, since structure formation is hierarchical and the structures hosting these high-redshift galaxies are therefore expected to be small. We, \textit{for the first time}, use the UV luminosity function data from JWST and combine them with the \textit{Planck} CMB data to probe neutrino self-interactions ($\nu$SI). If only one neutrino flavor  experiences such BSM interactions, then \textit{Planck} + JWST data give a 95\% C.L. upper limit on $\mathrm{Log}_{10}(G_{\mathrm{eff}}/\mathrm{MeV}^{-2})$ $\approx$ $-2.75$ and when all three flavors are interacting, the corresponding constraint becomes 
$\lesssim -3.41$; both
stronger than those obtained using only \textit{Planck} data.\,\,Our results highlight the power of this new observable, the high-redshift UV luminosity function from JWST as a complementary probe to terrestrial experiments, showing that small-scale observables can open a new window onto smaller neutrino self-interaction strengths.
\end{abstract}
	
\maketitle

\maketitle

\section{Introduction}
One of the most riveting endeavours of  current research community is to search for BSM physics. The discovery of neutrino masses provides one of the clearest glimpses of BSM physics~\cite{Giganti:2017fhf, SNO:2002tuh, Super-Kamiokande:1998kpq, KamLAND:2002uet}. Many extensions of the Standard Model (SM) that explain the origin of neutrino masses naturally predict the existence of new interactions involving neutrinos~\cite{Blinov:2019gcj, Berryman:2022hds, Singh:2026xak, Zhang:2026dcc, Perez-Castro:2026muj}. In the literature, BSM neutrino self-interactions ($\nu$SI) are often studied 
using an effective interaction of the form 
$\mathcal{L}_{\mathrm{eff}} = G_{\mathrm{eff}}(\bar{\nu}\nu)(\bar{\nu}\nu)$, 
where $G_{\mathrm{eff}}$ denotes the effective dimensionful coupling. Various 
astrophysical, cosmological, and laboratory observations probe values of 
$G_{\mathrm{eff}}$ that can be orders of magnitude larger than the Fermi 
constant $G_{\mathrm{F}}$ (= 1.16 $\times$ 10$^{-11}$ MeV$^{-2}$)~\cite{Cyr-Racine:2013jua, Oldengott:2014qra,Chacko:2003dt,Beacom:2004yd,Bell:2005dr,Cirelli:2006kt, Basboll:2008fx, Archidiacono:2013dua, Forastieri:2015paa, Forastieri:2017oma, Lancaster:2017ksf, Oldengott:2017fhy, Barenboim:2019tux, Foroughi-Abari:2026ofc, Laha:2013xua}.

\begin{figure}[htp!]
\includegraphics[width=1.0\columnwidth]{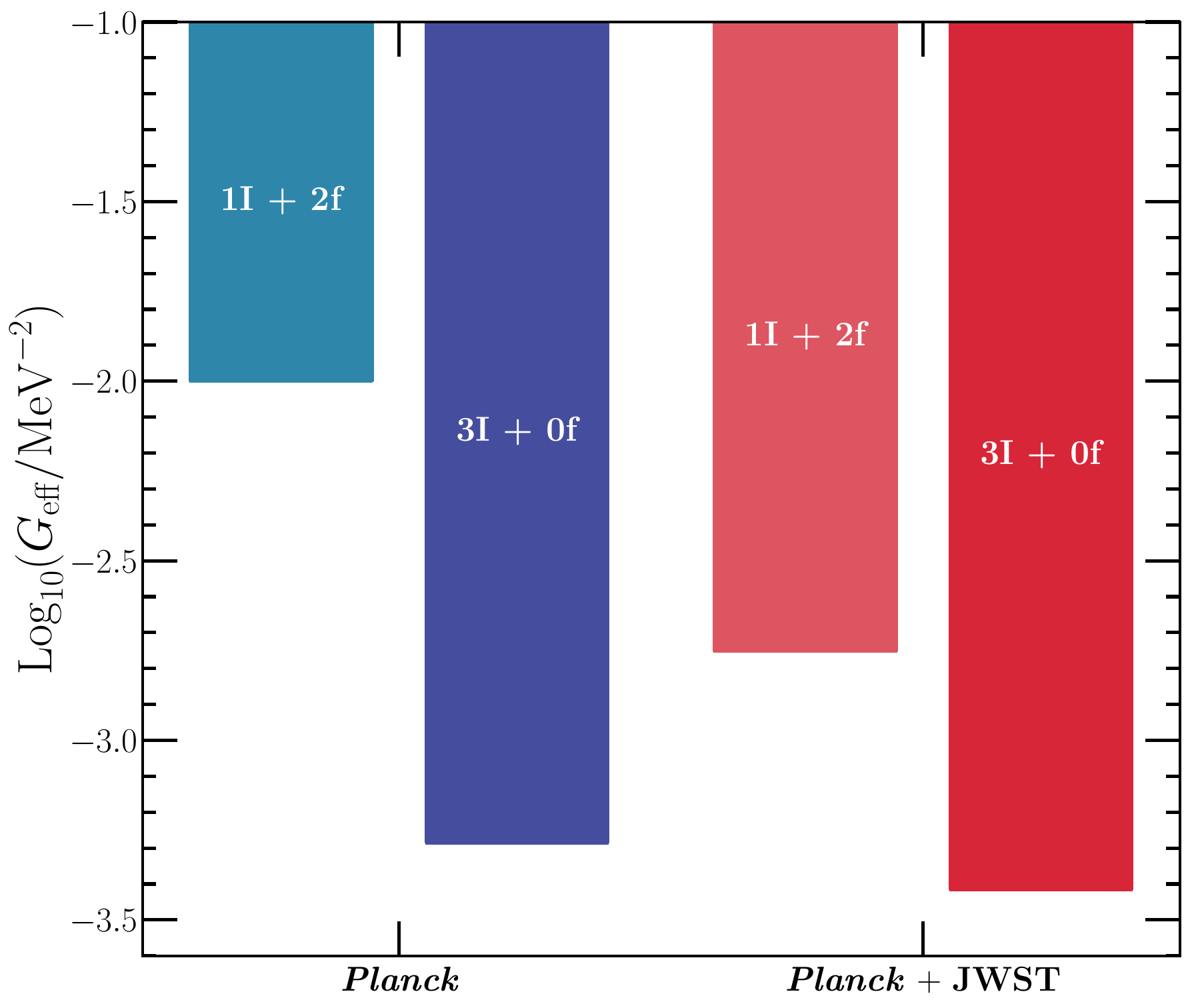}
\caption{Bar plot comparing the 95\% C.L. upper limits on the BSM neutrino self-interaction parameter $\mathrm{Log}_{10}(G_{\mathrm{eff}}/\mathrm{MeV}^{-2})$ obtained from the \textit{Planck}-only analysis (blue bars) and our analysis with the \textit{Planck}+JWST data (red bars) for both 1I +2f (lighter shade) and 3I +0f (darker shade) cases. We quote the upper limits in the \textit{End Matter}.}
    \label{fig:Constraint-Plot-Bar}
\end{figure}

\begin{figure*}[t]
\begin{center}
\includegraphics[width=\columnwidth]{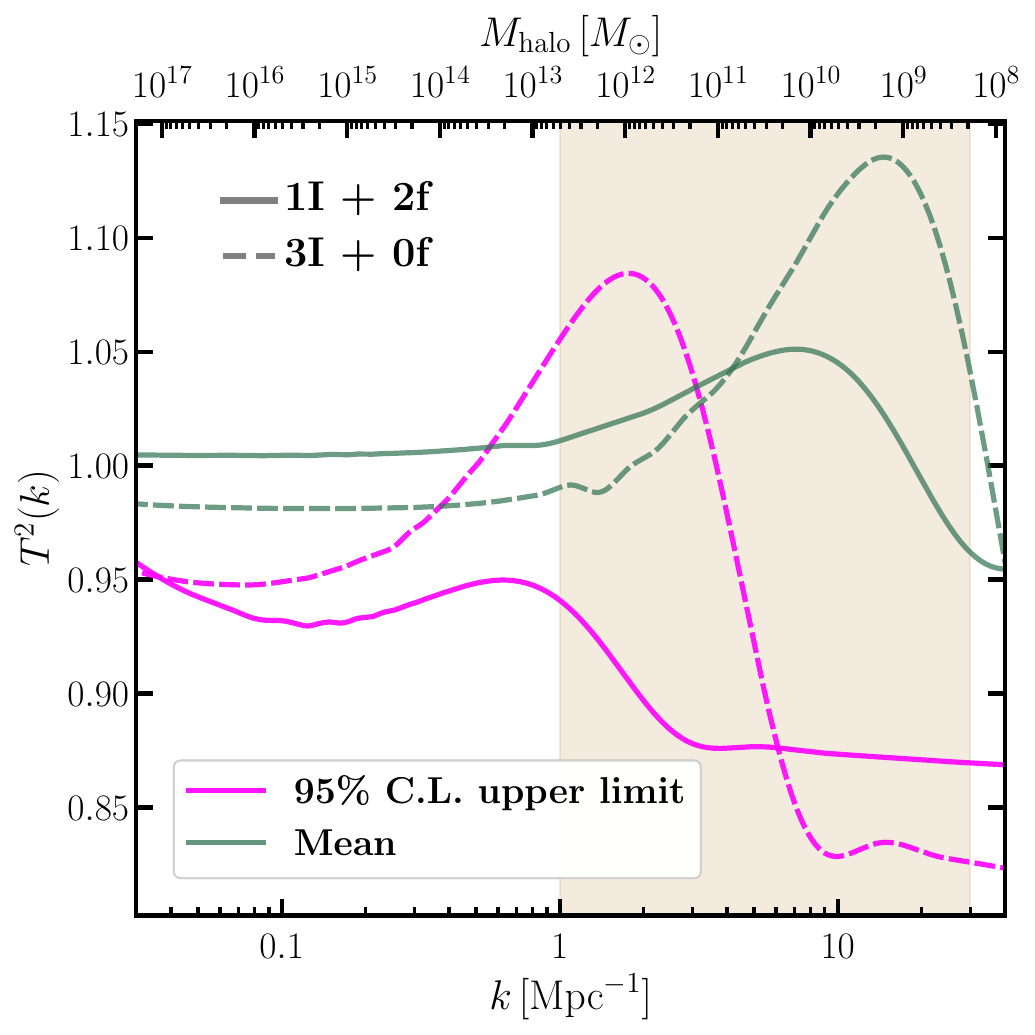}~~
\includegraphics[width=1.05\columnwidth]{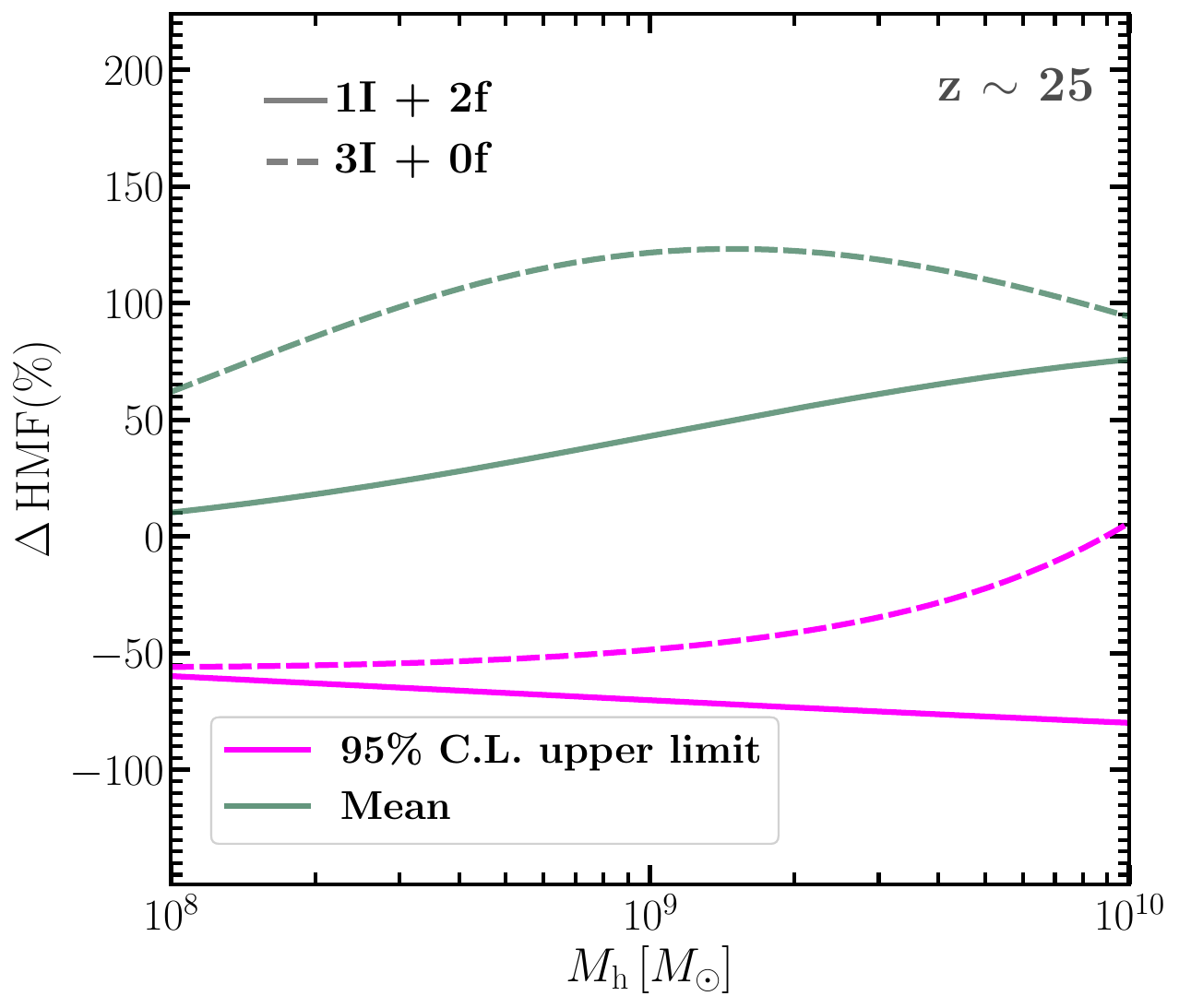}
\caption{\textbf{\textit{Left:}} Comparison of the ratio $T^{2}(k)=P_{\nu\mathrm{SI}, m_{\nu}}(k)/P_{\Lambda\mathrm{CDM},m_{\nu}}(k)$ of the matter power spectrum for $\nu\mathrm{SI} + \sum m_{\nu}$ to the baseline $\Lambda \mathrm{CDM}+\sum m_{\nu}$ model, for both flavor-specific (solid lines) and flavor-universal (dashed lines) cases. For all curves, the six $\Lambda$CDM parameters are fixed to the posterior mean values from the TT, TE, EE+lowE+lensing analysis in~\cite{Planck:2018vyg} and $\sum m_{\nu} = 0.06$. The green curves are evaluated using the mean values of $G_{\mathrm{eff}}$ and $\sum m_{\nu}$ from our posteriors with the \textit{Planck}+JWST likelihood, while the magenta curves are evaluated at the 95\% C.L. upper limits of these parameters. The yellow shaded region approximately indicates the range of the largest $k$-modes probed by JWST UVLFs~\cite{Winch:2024mrt}. \textbf{\textit{Right:}} Comparison of $\Delta \mathrm{HMF}(\%) = 100\times(\mathrm{HMF}_{\nu\mathrm{SI},m_{\nu}}-\mathrm{HMF}_{\mathrm{\Lambda CDM, m_{\nu}}})/\mathrm{HMF}_{\Lambda \mathrm{CDM}, m_{\nu}}$ at $z=25$, where $\mathrm{HMF}_{\nu\mathrm{SI},m_{\nu}}$ is the HMF calculated from $\nu\mathrm{SI}+\sum m_{\nu}$ cosmologies and $\mathrm{HMF}_{\Lambda\mathrm{CDM},m_{\nu}}$ is that calculated from the baseline $\Lambda\mathrm{CDM} + \sum m_{\nu}$ model. The color coding and line styles represent the same cases as in the transfer-function plot shown in the left panel.
    } 
		\label{fig:structure-formation}	
\end{center}
\end{figure*}

Our Universe can serve as a laboratory for BSM physics, providing complementary probes to terrestrial experiments.\,\,Neutrinos emitted from astrophysical sources can provide tests of $\nu$SI through their scattering with the low-energy relic neutrino background~\cite{Berryman:2022hds, Das:2022xsz, Kolb:1987qy, Farzan:2014gza, Shalgar:2019rqe}. These interactions can affect the physics of core-collapse supernova \cite{Shalgar:2019rqe,Bhattacharya:2023wzl, Dighe:2017sur}. The presence of $\sim  \mathrm{MeV}$--$\mathrm{EeV}$ astrophysical neutrinos~\cite{vanEeden:2026ojs, IceCube:2013cdw, IceCube:2013low, IceCube:2014stg, IceCube:2015gsk, IceCube:2018dnn, IceCube:2021rpz, IceCube:2023ame, LIGOScientific:2017ync, KM3NeT:2025npi,Baikal-GVD:2022fis, Dvornicky:2025kgo, Yuan:2025isv, Lazar:2026gde, Bai:2025pef, Verma:2026wrs} has made the study of $\nu$SI both timely and well-motivated~\cite{Hyde:2023eph, He:2025bex, Machado:2025ltu, BertolezMartinez:2025lyv,Franklin:2025kds,Foroughi-Abari:2025upe,Franklin:2024amy,Doring:2023vmk,Chang:2022aas,Akita:2022etk,Carpio:2022lqk,Babu:2019iml,Esteban:2021tub,Mazumdar:2020ibx,Chowdhury:2024fdr}.

Besides astrophysics, neutrinos from the early Universe can leave observable imprints on the history and evolution of the Universe~\cite{Dolgov:2002wy, Lesgourgues:2006nd, Hannestad:2010kz, Lesgourgues:2013sjj,Archidiacono:2017tlz,Giare:2025ath, Wang:2025ker, Racco:2024lbu,RoyChoudhury:2018gay,RoyChoudhury:2018vnm,RoyChoudhury:2018bsd,RoyChoudhury:2019hls,RoyChoudhury:2020dmd,MoradinezhadDizgah:2021upg,Sakstein:2019fmf,Chacko:2019nej,Corona:2021qxl,Bayer:2021kwg,Esteban:2021ozz,Yang:2020ope,Brandbyge:2010ge,Whitford:2024ecj}. Cosmology offers a complementary channel to probe several BSM scenarios, including $\nu$SI~\cite{Diacoumis:2018ezi, Mangano:2006mp, Wilkinson:2014ksa, DiValentino:2017oaw, Olivares-DelCampo:2017feq, Stadler:2019dii,Mosbech:2020ahp, Paul:2021ewd, Hooper:2021rjc, Mosbech:2022uud, Brax:2023tvn, Zhou:2026lqt, Dandoy:2025twl, Trojanowski:2025oro, Mosbech:2024wxr, Crumrine:2024sdn, Giare:2023qqn, Dey:2023sxx, Carenza:2022ngg, Li:2022dkc, Morgan:1981zy, Abazajian:2017tcc, DeGouvea:2019wpf, Das:2025asx, Bertolez-Martinez:2024wez,Ghosh:2024wva,Sharma:2024ctn,Sandner:2023ptm,Aloni:2023tff,Davoudiasl:2023uiq,Berbig:2023xqc,Das:2022xsz,Hagstotz:2020ukm,Chacko:2020hmh, EscuderoAbenza:2020cmq,Gehrlein:2019iwl,Mazumdar:2019tbm,Escudero:2019gvw,Ghosh:2019tab, Orlofsky:2021mmy, Dasgupta:2021ies,Kelly:2020aks,Barenboim:2020vrr,Froustey:2020mcq,Brandbyge:2010ge, Venzor:2020ova, Mogensen:2026vqj}. The delayed onset of neutrino free-streaming imprints characteristic signatures on cosmological observables, allowing $\nu$SI to be tested via Big Bang Nucleosynthesis (BBN), Cosmic Microwave Background (CMB), baryon acoustic oscillation (BAO), and large-scale structure (LSS) measurements~\cite{Kreisch:2019yzn, Perez-Castro:2026muj,Das:2023npl, GarciaEscudero:2025mtg, Whitford:2025dmq,Montefalcone:2025ibh, Poudou:2025qcx,Montefalcone:2025unv,Bostan:2023ped,Taule:2022jrz,Kreisch:2022zxp,Kumar:2022vee,Brinckmann:2020bcn,Das:2020xke, Kaplan:2024ydw,Pal:2024yom,Camarena:2023cku,He:2023oke,Pal:2026cgj,Parashari:2026dxo,Noriega:2025ulc,He:2025jwp, Venzor:2022hql,Forastieri:2019cuf, Libanore:2025ack, Whitford:2025dmq, Noriega:2026uhw}. The effect on structure formation is insensitive to the Dirac or Majorana nature of neutrinos.

For mediator masses $\mathcal{O}(\mathrm{MeV})$, initial analysis with \textit{Planck} CMB and BAO measurements favored two regions:\,\,the strongly interacting (SI) mode, $-2.5 \leq \mathrm{Log}_{10}(G_{\mathrm{eff}}/\mathrm{MeV}^{-2}) \leq 0.5$, and the moderately interacting (MI) mode, $\mathrm{Log}_{10}(G_{\mathrm{eff}}/\mathrm{MeV}^{-2}) \leq -2.5$~\cite{Kreisch:2019yzn, Lancaster:2017ksf, Das:2023npl, Libanore:2025ack}. While $\nu$SI models in the SI mode have been explored as a solution to the $H_0$ and $S_8$ tensions~\cite{Kreisch:2019yzn, Blinov:2019gcj, Berryman:2022hds}, they are strongly disfavored by terrestrial experiments~\cite{Brune:2018sab, Burgess:1992dt, Burgess:1993xh, KamLAND-Zen:2012uen, Agostini:2015nwa, Blum:2018ljv, Cepedello:2018zvr, Deppisch:2020sqh, Lessa:2007up, Bakhti:2017jhm, Ibe:2016dir, Dror:2020fbh, Ilten:2018crw, Brdar:2020nbj, Jones:2019tow, Bakhti:2018avv, Berryman:2019dme, Bardin:1970wq, Barger:1981vd, MINOS:2012ozn, Pasquini:2015fjv, FASER:2019dxq, FASER:2020gpr, SHiP:2020sos}. The SI mode has been ruled out by the more recent analysis with the Atacama Cosmology Telescope (ACT) DR6 dataset~\cite{AtacamaCosmologyTelescope:2025nti}. Beyond these interaction-strength regimes, more general scenarios can also be considered in which not all three neutrino flavors necessarily interact. We denote all three neutrino flavor interactions, two-flavor interactions + 1 free-streaming, and 1-flavor interaction + 2 free-streaming by 3I + 0f, 2I + 1f, and 1I + 2f, respectively. Since $\nu_{\tau}$ interactions are the least constrained~\cite{Blinov:2019gcj}, they leave open the possibility of using new observables to probe unexplored parameter regions.

\begin{figure}[t]
\begin{flushleft}
\includegraphics[width=1.0\columnwidth]{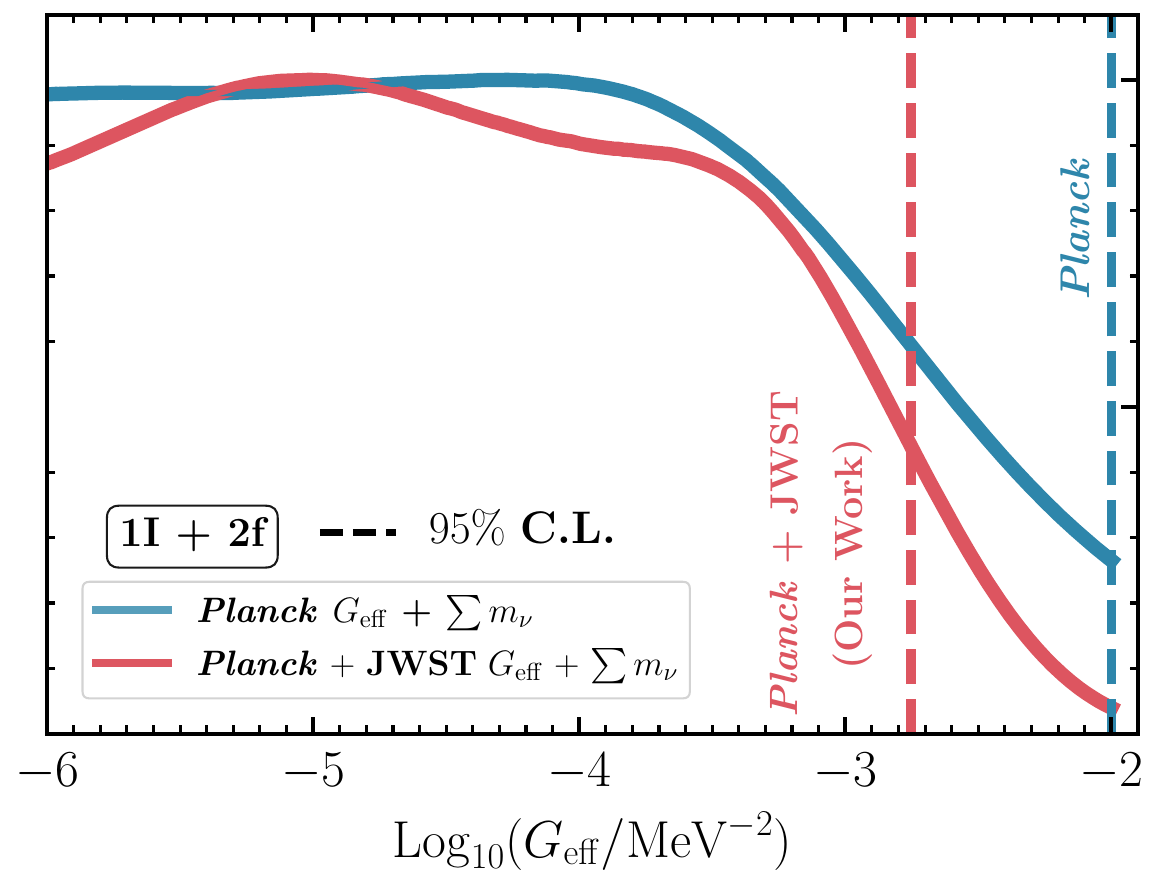}
\end{flushleft}
    \caption{\textit{\textbf{Left Panel:}} Comparison of the posterior distributions for the interaction strength $\mathrm{Log}_{10}(G_{\mathrm{eff}}/\mathrm{MeV}^{-2})$ for the $1\mathrm{I}+2\mathrm{f}$ coupling scenario, obtained from the \textit{Planck}-only analysis (blue solid line) and from our \textit{Planck}+JWST analysis (red solid line). The dashed vertical lines denote the corresponding 95\% C.L. upper limits on $\mathrm{Log}_{10}(G_{\mathrm{eff}}/\mathrm{MeV^{-2}})$ from \textit{Planck}-only analysis ($-2.00$) and from the combined analysis of \textit{Planck} + JWST ($-2.75$) in respective colors. 
}
    \label{fig:Geff-posterior-one-flavor}
\end{figure}

Observed galaxy populations from JWST have challenged our understanding of galaxy formation within the $\Lambda$CDM paradigm and pre-JWST astrophysics~\cite{2022ApJ...938L..15C,2022ApJ...940L..55F,2022arXiv220802794N,2023MNRAS.519.1201A,2023MNRAS.518.6011D,2023ApJS..265....5H,Yan:2022sxd,Robertson:2022gdk,2023MNRAS.518.4755A,2023MNRAS.523.1036B,2024ApJ...964...71H,2025Ap&SS.370...85H,2026MNRAS.548ag701R,2025arXiv250814972D,Matthee:2025fot,2026ApJ..1001...38T,2026MNRAS.545f2110J,2025ApJ...985...80R,Naidu:2025xfo,2025ApJ...983L..22K,2025ApJ...992...63W, 2024ApJ...960...56H, 2025ApJ...980..138H,2026ApJ..1002..134D,2026ApJ..1002..136W,2026PASA...43...37N,2026arXiv260416666M,2026ApJ..1009..134K, 2026A&A...708A..43K,2026ApJ...998L..36Z,2026MNRAS.546f2267C,2026ApJ...996..115A,2026OJAp....955261W}, opening the door to studying new astrophysical processes~\cite{2026ApJ...996..103K,Yung:2025ttv,Boylan-Kolchin:2022kae,Li:2023xky,2026ApJ...996..139S,2024ApJ...961...37C,2025MNRAS.540.3350A,2024A&A...689A.244C,Dekel:2023ddd,Fall:2009nf,2016MNRAS.455..334T,Yung:2023bng,Cueto:2023ayd,2024MNRAS.529.3563T,2024ApJ...967L..28M,Sun:2023ocn,2023MNRAS.526.2665S,2025ApJ...988L..10K,2024AJ....168..113C,2025ApJ...979..193C,2024MNRAS.533.1111E,2025A&A...697A..88L,2023MNRAS.522.6236T,2024ApJ...964..150D,2025NatAs...9..729H,Fakhry:2025yeu,Fakhry:2025tma,2025ApJ...982...23J,2025ScVar.155...95T,2025ApJ...978...89H,2025MNRAS.544.4390N,Binici:2024smk,Wang:2024piu,Matsumoto:2024eur,2024arXiv241011680D,Nusser:2024bcw,Ito:2024mla,Libanore:2023oxf,Chakraborty:2024xzn,2024MNRAS.531.2615C,Wang:2023gla,Glazebrook:2023vkx,2024ApJ...963...74W,Ventura:2024mbp,Qin:2023rtf,Pallottini:2023yqg,Wang:2023xmm,Gupta:2023mgg,Munoz:2026gxp,Bera:2025bik,Basu:2025jgs,Luberto:2025bct,Menon:2025yvv,Ferrara:2025rjc} and BSM theories~\cite{Klypin:2020tud,Shen:2024hpx,Biagetti:2022ode,Liu:2022bvr,Hutsi:2022fzw,Yuan:2023bvh,Dolgov:2023eqt,Parashari:2023cui,Gong:2022qjx,Du:2024afd,Nadler:2025crd,Urrutia:2025fvp,Kumar:2025gon,Dekker:2024nkb,Pilipenko:2024ywd,Menci:2024hop,Davari:2023tam,Tkachev:2024qua,Liu:2024yan,Winch:2023qzl,Wang:2023ros,Menci:2024rbq,Paraskevas:2023itu,Gouttenoire:2023nzr,Coulton:2024vot,Liu:2024edl,Paraskevas:2024ytz,Colazo:2024jmz,Forconi:2023hsj,Pilipenko:2024kuq,Hirano:2015wla,Lin:2023ewc,Wang:2023len,Adil:2023ara,Padmanabhan:2023esp,Comini:2026nsj,Acharya:2026wph,Blamart:2025szc,Sabti:2023xwo,Co:2025lrd,Shen:2025hat,Huang:2024aog,Prole:2025sif, Das:2025bnr}. These observations become sensitive to the microphysics that modifies structure formation at $k \sim \mathcal{O}(10)\,\mathrm{Mpc^{-1}}$~\cite{Parashari:2023cui, Hirano:2015wla, Sabti:2023xwo, Winch:2024mrt, Dayal:2023nwi, Liu:2024edl, Gilman:2021gkj,Das:2025bnr}.  In the MI regime, a wide range of $G_{\mathrm{eff}}$ values can modify the matter power spectrum $P(k)$, at these scales. JWST measures the comoving number density of galaxies within a given brightness range, making it an excellent probe of these interactions.

We present, {\it for the first time}, constraints on $\mathrm{Log}_{10}(G_{\mathrm{eff}}/\mathrm{MeV}^{-2})$ using JWST UV luminosity function (UVLF) data combined with the \textit{Planck} likelihoods. We discuss our data and likelihoods in the \textit{End Matter}. We restrict our analysis to the MI region,  for both flavor-universal (3I + 0f) and flavor-specific (1I + 2f) couplings. Including \textit{Planck} data with JWST improves sensitivity to the $\Lambda$CDM parameters and provides an independent cosmological probe, thereby breaking the degeneracy between cosmology and astrophysics.

Fig.~\ref{fig:Constraint-Plot-Bar} compares the 95\% C.L. upper limits on $\mathrm{Log}_{10}(G_{\mathrm{eff}}/\mathrm{MeV}^{-2})$ from \textit{Planck}-only data (blue shades) and \textit{Planck} + JWST data (red shades). For both 1I + 2f and 3I + 0f cases, the combined likelihoods give better constraints on $G_{\mathrm{eff}}$ compared to \textit{Planck} only analysis. Our work shows that JWST UVLFs are sensitive to suppression in $P(k)$ on scales $k \sim \mathcal{O}(10)\,{\rm Mpc}^{-1}$, through their impact on the halo mass function (HMF) at high redshifts. Consequently, these data can constrain $G_{\mathrm{eff}}$ that suppress $P(k)$ around those scales.

\section{Effect on the structure formation}
We briefly discuss the impact of $\nu$SI on $P(k)$~\cite{Kreisch:2019yzn, Camarena:2023cku, Libanore:2025ack, Pal:2024yom, Pal:2026cgj, He:2023oke}. For both 3I+0f and 1I+2f cases, the delay in neutrino free streaming reduces the anisotropic stress compared to $\Lambda$CDM cosmologies. The onset of free streaming occurs at approximately the same time in both the flavor-specific and flavor-universal cases. This suppression in the anisotropic stress results in nontrivial changes in the evolution of the gravitational potential $\psi$, which drives the growth of matter perturbations~\cite{Kreisch:2019yzn} 

\begin{equation}
\ddot{d}_{c} + \frac{\dot{a}}{a}\dot{d}_{c} = -k^{2}\psi \,,
\label{eq:fluctuation-growth}
\end{equation}
where $d_{c}$ is the dark matter (DM) density contrast in the Newtonian gauge, $a$ is the scale factor, and $k$ is the comoving wavenumber of the perturbation mode.\,The dots in Eq.~\eqref{eq:fluctuation-growth} denote derivatives w.r.t the conformal time.

Every value of $G_{\mathrm{eff}}$ can be associated with a scale, $k_{\mathrm{fs}}$,  that enters the horizon during the onset of free-streaming. Modes with $k < k_{\mathrm{fs}}$, that enter the horizon well after decoupling, follow the standard $\Lambda$CDM evolution. Modes entering the horizon around the onset of free streaming, $k \simeq k_{\mathrm{fs}}$, receive a boost at horizon entry but quickly converge to the standard $\Lambda$CDM evolution, causing an enhancement in the matter power spectrum. Modes with $k > k_{\mathrm{fs}}$, which enter the horizon when neutrinos remain tightly coupled, also receive this boost at horizon entry, but eventually experience an overall suppression. This imprints a scale-dependent enhancement and suppression in $P(k)$ for $\nu\mathrm{SI}$ cosmologies (see left panel of Fig.~\ref{fig:structure-formation}). We solve the modified cosmological perturbation equations for $\nu$SI using \texttt{nuCLASS}\footnote{\url{https://github.com/subhajitghosh-phy/nuCLASS/tree/nuself}} to obtain $P(k)$.

In the left panel of Fig.~\ref{fig:structure-formation}, we show the impact of $\nu\mathrm{SI}$ and $\sum m_{\nu}$ on the matter power spectrum $P_{\nu\mathrm{SI},\,m_{\nu}(k)}$ using $T^{2}(k) = P_{\nu\mathrm{SI},\,m_{\nu}}(k)/P_{\Lambda\mathrm{CDM},\,m_{\nu}}(k)$, where $P_{\Lambda\mathrm{CDM},\,m_{\nu}}(k)$ is computed for a baseline $\Lambda\mathrm{CDM}+\sum m_\nu$ model with six $\Lambda$CDM parameters fixed to the Planck TT, TE, EE + lowE + lensing bestfit values from~\cite{Planck:2018vyg} and $\sum m_\nu=0.06\,\mathrm{eV}$. The green curves, evaluated at the mean values of $G_{\mathrm{eff}}$ and $\sum m_\nu$, show an enhancement near the largest $k$-scales probed by JWST, indicated by the yellow band. The magenta curves, evaluated at 95\% upper limits, instead show a suppression in both the 1I + 2f and 3I + 0f cases, as shown by the solid and dashed lines, respectively. The upper horizontal axis in the left panel shows the corresponding approximate halo mass $M_{h}(k) = (4\pi/3)\rho_{m,0}(\pi/k)^{3}$~\cite{Nadler:2025fcv, Nadler:2019zrb}, where $\rho_{m,0}$ is the present day mean matter density.

\section{Model for the UV luminosity function}\label{sec:UVLF}

Since JWST observes UV bright galaxies, we use the observable UVLF, which tracks the number densities of galaxies per unit UV magnitude. Since the UVLF traces the abundance of luminous galaxies, which are assumed to form within DM halos, it is sensitive to structure formation and hence to $P(k)$. One also needs to model astrophysics to understand galaxy formation within DM halos. Modelling the UVLF has two components: 1) the HMF and 2) galaxy--halo connection prescription. We calculate the UVLF from $P(k)$ following\,\cite{Das:2025bnr}.

To calculate the HMF, we adopt ``smooth-$k$'' window function with parameters chosen from \cite{Leo:2018odn, Verwohlt:2024efh}, and the Sheth--Tormen fitting function~\cite{Sheth:1999su}. The effect of $\nu$SI enters through $P(k)$ via mass variance at redshift $z$, $\sigma^{2}(M_{h},z)$, where $M_{h}$ is the halo mass:
\begin{equation}
\sigma^{2}(M_{h},z) = \frac{1}{2\pi^{2}} \int_{0}^{\infty} P(k, z)\,W^{2}(kR)\,k^{2}\,dk ,
\end{equation} where $W(kR)$ is the window function. We use the galaxy--halo connection prescription from\,\cite{Das:2025bnr}, with the star formation efficiency (SFE) calibrated from \textsc{Thesan-Zoom} simulations  \footnote{\url{https://www.thesan-project.com/thesan-zoom/movies.html}}~\cite{Shen:2025isu, Shen:2025isu,Kannan:2025ohh,2025MNRAS.544..513M,Zier:2025oqj,2025MNRAS.544..410Z,2025MNRAS.544.1732M,Wang:2025oqg,2026MNRAS.545f2092M,2026MNRAS.548ag016M,2026MNRAS.546ag064P,2026ApJ..1001..188G,2025arXiv251013952M}. We use empirical relations between the specific UV luminosities of galaxies and the star formation rate (SFR) to calculate the absolute UV magnitude $M_{\mathrm{UV}}$ as a function of $M_{h}$. However, this relation is not deterministic; it shows some scatter. To model this burstiness, we adapt the same formalism as \cite{Das:2025bnr} adapted from~\cite{Shen:2025isu, Sabti:2021xvh, Shen:2024hpx}. We calculate the UVLF at a redshift z, $\bar{\Phi}_{\rm UV}^{i}$, by 
\begin{equation}
\bar{\Phi}^i_{\mathrm{\rm UV}}(z)
= \frac{1}{\Delta M^{i}_{\mathrm{\rm UV}}}
\int \mathrm{d}M'_{\mathrm{\rm UV}}
\int \mathrm{d}M_h \frac{\mathrm{d}n (z)}{\mathrm{d}M_h}\,
p(M'_{\mathrm{\rm UV}}| M_{h})\,,
\label{eq:UVLF_binned}
\end{equation}
where $\Delta M_{\rm UV}^{i}$ is the width of the $i$-th UV-magnitude bin. Here $n$ is the comoving number density of halos, $dn(z)/dM_{h}$ is the HMF at redshift z, and $p(M_{\mathrm{UV}}|M_{h})$ is the probability that a DM halo of host mass $M_{h}$ can host a galaxy with UV magnitude $M_{\mathrm{UV}}$. The outer integral in eq.~\eqref{eq:UVLF_binned} is over the $i$-th magnitude bin.  

\section{Results}
\label{sec:results}
We sample the cosmological and nuisance parameters using the \textit{Planck} 2018 high-$\ell$ TT, TE, EE, low-$\ell$ TT and EE likelihoods, the \textsc{Pantheon} supernova likelihood, and JWST UVLFs implemented through a modified \textsc{Gallumi} pipeline. The sampling is performed with \textsc{MontePython}~\cite{2013JCAP...02..001A, Speagle:2019ffr}, a Monte Carlo Markov Chain (MCMC) package, and our Bayesian setup is described in the \textit{End Matter}. All main-text results are obtained in an extended $\Lambda$CDM model where we vary the six $\Lambda$CDM parameters ${\omega_{b}, \omega_{\mathrm{cdm}}, 100\theta_{s}, A_{s}, n_{s}, \tau_{\rm reio}}$~\cite{Planck:2018vyg}, together with $\sum m_{\nu}$ and $\mathrm{Log}_{10}(G_{\mathrm{eff}}/\mathrm{MeV}^{-2})$. When marginalizing over $\sum m_{\nu}$, we set the lower prior boundary to $0.06\,\mathrm{eV}$, motivated by the minimum value of $\sum m_{\nu}$ allowed by the normal mass ordering. If we choose the lower prior boundary at inverted neutrino mass ordering $\sum m_{\nu} = 0.1\,\mathrm{eV}$, the constraints would become stronger. This is because a larger value of $\sum m_{\nu}$ leads to stronger suppression of structure formation, thereby increasing the sensitivity to smaller interaction strengths. Larger cross sections induce greater suppression at larger $k$ scales and are therefore increasingly disapproved by the data.

Our UVLF likelihood includes only high-redshift data ($z \gtrsim 9$), when the Universe was still in its infancy. In $\Lambda$CDM structure grows hierarchically, so the DM halos expected at these redshifts are smaller~\cite{2025ScVar.155...95T}, and correspond to $k \sim 1 $--$ 10\,\mathrm{Mpc}^{-1}$. In Fig.~\ref{fig:structure-formation}, the $T^{2}(k)$ curves evaluated at 95\% C.L. upper limits of $G_{\mathrm{eff}}$ and $\sum m_{\nu}$ show clear suppression at these scales relative to the curves evaluated at their posterior mean values from their posteriors, suggesting that the UVLF data disfavour parameters that suppress structure formation on JWST-relevant $k$-scales. This inference is consistent with the observed increase in the population of high-redshift structures relative to pre-JWST predictions.

In the right panel of Fig.~\ref{fig:structure-formation}, we show the percentage change in the HMF due to $\nu$SI, $\Delta\mathrm{HMF}(\%)$. We present $\Delta\mathrm{HMF}$ because it provides a cleaner probe of the cosmological effects of the new physics. Unlike the UVLF, which depends on the interplay between astrophysics and cosmology, the HMF is more directly tied to cosmology. At $z=25$, we find $\Delta\mathrm{HMF}(\%)<0$ for the 95\% C.L. upper limits values of $\sum m_{\nu}$ and $G_{\mathrm{eff}}$, indicating suppressed structure formation relative to the HMF evaluated at the posterior-mean parameter values, which instead shows an enhancement at $M_h\sim10^{8}$--$10^{10}\,M_{\odot}$, the mass scales relevant at this redshift. Although $\Delta\mathrm{HMF}(\%)$ follows a similar trend to the transfer function over $k\sim1$--$10\,\mathrm{Mpc}^{-1}$, a one-to-one mapping between $P(k)$ and the HMF can be misleading, since the HMF depends on $P(k)$ through $\sigma(M_h)$ and its derivative (see \textit{Supplemental Material}).

The improvement in $G_{\mathrm{eff}}$ constraints is mainly driven by the UVLFs in the high-redshift bins, as they probe larger wavenumbers $k$. Since structure formation is already less efficient at higher redshifts even in $\Lambda$CDM, the presence of galaxy candidates at such early times makes the data more sensitive to any additional suppression relative to $\Lambda$CDM. We also show $\Delta\mathrm{HMF}(\%)$ for the $z=9$ and $z=14$ bins in the \textit{Supplemental Material}. Even at these redshifts, the HMFs evaluated at the 95\% C.L. upper limits of $G_{\mathrm{eff}}$ and $\sum m_{\nu}$ show suppression, while those evaluated at the posterior mean values show enhancement. This suggests that the constraints are mainly driven by the suppression of $P(k)$, and consequently of the HMF, in the $\nu$SI models, which is disfavored by the data across different redshifts.

In Fig.~\ref{fig:Geff-posterior-one-flavor}, we compare the 1D marginalized posteriors of $\mathrm{Log}_{10}(G_{\mathrm{eff}}/\mathrm{MeV}^{-2})$ for the 1I + 2f case from \textit{Planck}-only (blue lines) and \textit{Planck}+JWST (red lines) analyses. The colored dashed lines show the 95\% C.L. upper limits for the respective cases. We find that $\mathrm{Log}_{10}(G_{\mathrm{eff}}/\mathrm{MeV}^{-2})$ is more tightly constrained when JWST is included. For the posterior derived using \textit{Planck} data, the 95\% C.L. upper limits on $\mathrm{Log}_{10}(G_{\mathrm{eff}}/\mathrm{MeV}^{-2}) \sim -2$ lies at the upper prior boundary. This is due to the fact that the \textit{Planck} data have very weak sensitivity for the one-flavor interacting case. For structure formation, the effect is reduced to approximately one-third of that in the three-flavor interacting case. This weakens the constraint, since larger values of $G_{\rm eff}$ are required to produce a comparable suppression on JWST-relevant scales. However, the UVLF is highly sensitive to changes in $\sigma(M_h)$, allowing the effect to remain observable even when its magnitude is reduced. This provides JWST with additional sensitivity to $G_{\rm eff}$ that is substantially weaker in \textit{Planck} alone. We performed the same analysis using a ``\textit{top-hat}" window function~\cite{Parashari:2023cui}. For both the one-flavor interacting and flavor-universal cases, the 95\% C.L. upper limits change by only $3$--$4\%$, owing to the large observational uncertainties and marginalization over the astrophysical parameters.

\section{Discussions and Conclusions}
\textit{For the first time}, we probe $\nu\mathrm{SI}$, using JWST UVLF measurements  in combination with the \textit{Planck}-\texttt{lite} likelihoods.\,We find that, for both flavor-specific and flavor-universal couplings, the \textit{Planck}+JWST likelihood yields stronger constraints on $G_{\mathrm{eff}}$ than the \textit{Planck}-only analysis, because of the sensitivity of the UVLF to the change in structure formation due to $\nu$SI. Furthermore, we show that for a general class of 1I + 2f scenarios, \textit{Planck}+JWST data can give competitive upper limits. 

Cosmology provides the strongest constraints on $\sum m_{\nu}$ through its effects on geometry and structure growth~\cite{Herold:2024nvk,Lesgourgues:2012uu, Archidiacono:2017tlz, Loverde:2024nfi, Wong:2011ip, Archidiacono:2016lnv, Lesgourgues:2006nd, Zhou:2025nkb}. Increasing $\sum m_{\nu}$ suppresses $P(k)$ above $k \sim k_{\mathrm{nr}}$, modes that enter the horizon when neutrinos become non-relativistic. Since JWST indirectly probes the matter power spectrum, we vary $\sum m_{\nu}$ in our MCMC analysis and marginalize over it. We show the $\sum m_{\nu}$ posteriors in the \textit{Supplemental Material}.

Unlike warm dark matter like scenarios~\cite{Eberhardt:2025caq,Dekker:2021scf,Nadler:2025fcv}, $\nu\mathrm{SI}$ does not introduce a cutoff scale in $P(k)$, but instead produces characteristic enhancement and suppression at different $k$ values. Although the induced changes are $\lesssim 15\%$, these distinctive simultaneous effects on both large and small scales indicate that a joint analysis probing $P(k)$ across a broad range of $k$ can be particularly advantageous for constraining $\nu$SI. Using observables such as LSS, Milky Way satellite counts, the subhalo mass function, Lyman-$\alpha$, weak lensing, 21-cm observations, and UVLFs could provide stronger and more robust limits on $G_{\mathrm{eff}}$. 

$\nu$SI can also be probed using Hubble Space Telescope (HST) UVLF data, which have significantly smaller error bars. We focus on JWST because it provides the highest-redshift data. We do not combine HST and JWST, since astrophysical modelling across different redshifts can differ significantly~\cite{2026ApJ...996..103K}. In addition to \textsc{Thesan-Zoom}, several other simulations have carefully modelled the astrophysics at those high redshifts~\cite{2026arXiv260526211B, Samuel:2026ymj, 2026arXiv260506782L, 2026ApJ..1009..134K}. Incorporating a more comprehensive astrophysical framework would make our constraints even more robust.

JWST has been collecting data for only four years and has already made a major impact. With more observations, it is expected to discover more galaxies and reduce the observational uncertainties. However, estimating its future sensitivity requires a Fisher forecast analysis. Instead, we take a more physical approach by identifying the values of $G_{\mathrm{eff}}$ for which the effects of $\nu$SI extend to the largest wavenumbers probed by JWST. This provides an estimate of the potential reach of JWST in both flavor-universal and flavor-non-universal scenarios, as discussed in the \textit{Supplemental Material}. With the current error bars, our analysis for the 3-flavor interacting case provides weaker constraints than other probes, such as the Lyman-$\alpha$ forest~\cite{He:2025jwp}; however, with more data, JWST could become comparable to or even more sensitive than these probes in constraining such interactions.

This is the first time UVLFs have been used to probe $\nu$SI. High-redshift UVLF data are among the most powerful probes of small-scale structure~\cite{Gilman:2021gkj, Lazare:2025gha, Das:2025bnr}. With its unprecedented technology, JWST promises to offer a complementary window onto BSM physics by unfolding the formation of the first structures in our Universe.

\label{sec:Conclusion}
\begin{acknowledgments}
\section{Acknowledgments}
We especially thank Subhajit Ghosh for insightful discussions throughout the project.\,\,R.M. is thankful to Abhijeet Singh for implementing the smooth-k filter function in the $\texttt{CLASS}$ code. R.M especially acknowledges Debajit Bose, Eung Jin Chun, Nirmal Raj, Sudhir Vempati, and Sudip Jana for their discussions on the particle physics models.\,\,R.M is thankful to Ujjwal Kumar Upadhyay and Yashi Tiwari for discussions on \textsc{MontePython} MCMC analysis.\,\,We acknowledge Shadab Alam, Susmita Adhikari, Shamim Akhtar, Carlos Argüelles-Delgado, Hakim Atek, Arka Banerjee, Indra Banerjee, Jeet Bhattacharya, Subhadip Bouri, Denise Burgarella, Anirban Chakraborty, Sanskrita Chakraborty, Suchetana Chatterjee, Pravabati Chingangbam, Shouvik Roy Chowdhury, Debtosh Chowdhury, Deep Jyoti Das, Souradeep Das, Basudeb Dasgupta, Bihag Dave, Amol Dighe, Mansi Dhuria, Jaya Doliya, Abhishek Dubey, Raj Gandhi, Deep Ghosh, Durba Ghosh, Srubabati Goswami, Shunshaku Horiuchi, Sahabub Jahedi, Seth Koren, Tarak Nath Maity, Supratik Pal, Maxim Perelstein, Santosh Kumar Rai, Mayuri S. Rao, Namrata Roy, Tirthankar Roychowdhury, Akash Kumar Saha, Hamsini S., Anuvab Sarkar, Tanoy Sarkar, Martin Schmaltz, Manibrata Sen, Rinchen Sherpa, Aadarsh Singh, Tarun Souradeep, Pushkar Tripathi, Jessica Turner, Ken Van Tilburg, Amol Pattawardhan, Tracy Slatyer, Maria Volpe, and Yvonne Y. Y. Wong for discussions and comments. We also acknowledge the use of ChatGPT, Claude, and Perplexity as AI-assisted tools during the preparation of this manuscript, including for language editing, literature searches, and assistance with code development and debugging. R.M. is supported by the ISRO-IISc STC Grant No. ISTC/PHY/RL/499 and ANRF National Postdoctoral Fellowship (NPDF) under Project No. PDF/2025/001161. AD acknowledges ANRF for the financial support through PMECRG (Grant no. ANRF/ECRG/2025/001012/PMS). VG was supported by the John Templeton
Foundation Grant No.
63667, the Research Corporation for Science Advancement Cottrell
Scholar Program, the IBM Einstein Fellowship at the Institute for
Advanced Study, and the NSF
CAREER Grant No. PHY-2239205. \,\,R.L. acknowledges financial support from the institute start-up funds and ANRF for the grant
no. ANRF/ARG/2025/005140/PS. We would like to thank Vinayak Sanil, who maintained the computing workstations, which were invaluable for various computations performed in this work. Also, he uploaded the short video explaining the work on YouTube~\footnote{\href{https://youtu.be/gM9Zw2QHYcE}{https://youtu.be/gM9Zw2QHYcE}}. R. L. would like to acknowledge support from the ICTP through the Associates Programme and from the Simons Foundation, whose grant
(Record ID: SFI-MPS-T-Institutes-00012057, AD) also
supported this work. 
\end{acknowledgments}

\newpage


\nocite{Boddy2022, An:2024nsw, Ma1995}
\nocite{Yin:2018yjn, CRESST:2019jnq, CRESST:2019axx, CRESST:2017ues, EDELWEISS:2019vjv, Ema:2018bih, Emken:2021lgc, Sabti:2019mhn, Erickcek:2007jv, Mahdawi:2018euy, Essig:2012yx, Essig:2017kqs, XENON10:2011prx, XENON:2016jmt, DarkSide:2018ppu, SuperCDMS:2018mne, Crisler:2018gci, SENSEI:2019ibb, SENSEI:2020dpa, Fortin:2011hv, Essig:2011nj, Chu:2011be, Lesgourges2011, Press1974, Murray2013, Sheth:1999mn, 2024MNRAS.530.4868Y, Madau2014}
\bibliographystyle{JHEP}
\bibliography{refs_nuSI.bib}

\clearpage

\section{End Matter}

\renewcommand{\theequation}{A\arabic{equation}} 
\setcounter{equation}{0} 
\renewcommand{\thetable}{A\arabic{table}} 
\setcounter{table}{0} 

\subsection{Data and Methodology}

The JWST UVLF measurements used for this analysis are taken from a set of 25 galaxies presented in~\cite{2024ApJ...960...56H}, 10 galaxies from~\cite{2025ApJ...980..138H}, 7 galaxies from~\cite{2025A&A...693A..50N}, 2 galaxies from~\cite{Naidu:2025xfo}, 2 galaxies from\,\cite{2025ApJ...983L..22K}, 5 galaxies from\,\cite{Castellano:2025vkt}, 9 galaxies from\,\cite{Perez-Gonzalez:2025bqr}, and upper limits on UVLFs from\,\cite{Castellano:2025vkt, 2025ApJ...992...63W, 2026ApJ..1002..136W}. We only consider UVLF data above redshift $z \gtrsim 9$. All data and upper/lower limits till redshift $\sim 16$ are \textit{spectroscopic} measurements, while we include very high-redshift photometric data, upper, and lower limits for $z \gtrsim 17$. At every redshift bin, we set the width of the brightness magnitude $M_{\mathrm{UV}}$ bin around every $M_{\mathrm{UV}}$ points either to the width given in the relevant paper, or by selecting the larger of the differences to the adjacent $M_{\mathrm{UV}}$ values. This choice of magnitude binning is the same conservative approach as adapted in~\cite{Das:2025bnr}.

\subsection{Likelihood Analysis}
We use the following likelihoods to derive our results:
\begin{itemize}
    \item \textit{\textbf{Planck:}} We use the \textit{Planck} 2018 \texttt{plik\_lite} high-$\ell$ TT/TE/EE likelihood, along with the \texttt{commander} low-$\ell$ TT likelihood and the \texttt{SimAll} low-$\ell$ EE likelihood.
    \item \textbf{JWST:} We modified the \textsc{Gallumi} pipeline for HST~\cite{Sabti:2021xvh} in the same way as described in~\cite{Das:2025bnr}, using an astrophysical formalism motivated by high-redshift zoom-in simulations. The likelihood function chosen for the $M_{\mathrm{UV}}$-binned $\Phi_{\mathrm{UV}}$ is
\begin{equation}
\log \mathcal{L}
=-\frac{1}{2}\sum_{z}\sum_{i}
\begin{cases}
\qty(\dfrac{\bar{\Phi}^i_{\rm UV}-\Phi^i_{\rm UV}}
{\delta\Phi_{\rm UV}^{{\rm up}, i}})^{2},
& \bar{\Phi}^i_{\rm UV}>\Phi^i_{\rm UV}\\
\qty(\dfrac{\bar{\Phi}^i_{\rm UV}-\Phi^i_{\rm UV}}
{\delta\Phi_{\rm UV}^{{\rm lo}, i}})^{2},
&  \bar{\Phi}^i_{\rm UV}<\Phi^i_{\rm UV}
\end{cases}
\end{equation}
where $\bar{\Phi}_{\rm UV}^i$ is the UVLF calculated in the $i^{\rm th}$ magnitude bin using Eqn.~\eqref{eq:UVLF_binned}. The JWST data in the $i^{\rm th}$ $M_{\rm UV}$ bin is given by
\begin{equation}
\qty(\Phi_{\rm UV}^i)^{+\delta\Phi_{\rm UV}^{{\rm up},i}}_{-\delta\Phi_{\rm UV}^{{\rm lo},i}} .
\end{equation}
The redshift dependence of our model enters through the HMF. The likelihood function defined in our code can accommodate asymmetric error bars. We incorporate dust corrections following an extrapolation of Ref.~\cite{Sabti:2021xvh} to redshifts $z \approx 10$, which closely matches the \textsc{Thesan-Zoom} simulations\,\cite{Kannan:2025ohh}.
\item \textbf{Pantheon:} We used the distance moduli measurements from the Pantheon Type-Ia supernovae likelihood (SNe) catalogue, following~\cite{Sabti:2021xvh}.
\end{itemize}
\section{Parameters \& Priors}
We perform the likelihood analysis using \textsc{MontePython} for both flavour-universal and flavour-specific couplings. In each case, we vary the six standard cosmological parameters, $\{\omega_{b}, \omega_{\mathrm{cdm}}, 100\,\theta_{s}, A_{s}, n_{s}, \tau_{\rm reio}\}$, together with the neutrino mass parameter, $\sum m_{\nu}$, and the logarithm of the neutrino self-interaction strength, $\mathrm{Log}_{10}(G_{\rm eff}/\mathrm{MeV^{-2}})$.

The nuisance parameter for the \textit{Planck} likelihood is the lensing amplitude $A_{\rm lens}$. For the JWST likelihood, there are several astrophysical nuisance parameters:\,\,$\{\alpha_\star, \beta_\star, \mathrm{Log}_{10}\epsilon_0, \mathrm{Log}_{10}M_0\},$ which characterize the SFE, along with the normalization of the UV variability $\sigma_{\rm UV}(10^{10.5}M_\odot)$. For the Pantheon likelihood, the nuisance parameter is the intrinsic SN Ia magnitude $M$.

Despite using $\textit{Planck}$ likelihoods, we impose Gaussian priors on $A_s$ and $n_s$ from CMB measurements in our JWST likelihoods to reduce the residual degeneracy between $A_s$, $n_s$, and $G_{\rm eff}$. Since JWST is sensitive to small-scale structure formation, variations in $A_s$ and $n_s$ can broaden the inferred constraint on $\mathrm{G}_{\rm eff}$. These priors, therefore, help stabilize the analysis, rather than provide independent information. We also impose Gaussian priors on $\omega_{b}$ from BBN~\cite{2009PhR...472....1I}. We assume uniform priors for the remaining parameters, including $\sum m_{\nu}$ and $\mathrm{Log}_{10}(G_{\mathrm{eff}}/\mathrm{MeV}^{-2})$. We present the likelihoods and priors in Tables~\ref{tab:likelihood_summary} and \ref{tab:prior_summary}. The table Tab.~\ref{tab:constraints_summary_combined} show the mean the and 95\% upper limits on $G_{\mathrm{eff}}$ and $\sum m_{\nu}$ from the \textit{Planck}-only analysis and the \textit{Planck} + JWST analysis.

We use \textsc{MontePython} to perform MCMC analysis with the likelihoods and priors discussed in the preceding paragraphs. At the end of each run, we ensure that the Gelman--Rubin criterion for every parameter satisfies $R \lesssim 1.05$. To obtain the 95\% C.L. upper limits on $\mathrm{Log}_{10}(G_{\mathrm{eff}}/\mathrm{MeV}^{-2}),$ we marginalize over the posterior samples. We use \texttt{GetDist}~\cite{Lewis:2019xzd} for our statistical analysis and plotting.  

\begin{table*}[t]
\centering

\begin{tabular}{l@{\hspace{1.2cm}}l}
\hline\hline
\textbf{Shorthand} & \textbf{Likelihoods} \\
\hline\hline
\textit{Planck} &
\begin{tabular}[t]{@{}l@{}}
\textit{Planck} 2018 high-$\ell$ TT, TE, EE \\
+ low-$\ell$ TT, EE
\end{tabular}
\\
\hline
\textit{Planck} + JWST &
\begin{tabular}[t]{@{}l@{}}
\textit{Planck} + JWST UVLF \\
+ BBN $\omega_b$ prior \\
+ CMB $A_s$ and $n_s$ prior
\end{tabular}
\\
\hline\hline
\end{tabular}

\caption{Summary of the different likelihood combinations and priors used in our analysis.}
\label{tab:likelihood_summary}

\vspace{0.3cm}

\begin{tabular}{l@{\hspace{1.2cm}}l}
\hline\hline
\textbf{Parameter} & \textbf{Prior range} \\
\hline\hline
$100\omega_b$ &
$\mathcal{N}(2.2377,0.015)$ \\
$\omega_{\mathrm{cdm}}$ &
$\mathcal{U}[-\infty,\infty]$ \\
$A_s\times10^9$ &
$\mathcal{N}(2.1,0.03)$ \\
$n_s$ &
$\mathcal{N}(0.9649,0.0042)$ \\
$100\theta_s$ &
$\mathcal{U}[-\infty,\infty]$ \\
$\tau_{\rm reio}$ &
$\mathcal{U}[-\infty,\infty]$ \\
$\sum m_\nu\,[\mathrm{eV}]$ &
$\mathcal{U}[0.06,1]$ \\
$\mathrm{Log}_{10}(G_{\mathrm{eff}}/\mathrm{MeV}^{-2})$ &
$\mathcal{U}[-6,-2]$ \\
\hline
$\alpha_\star$ &
$\mathcal{U}[0.0,3.0]$ \\
$\beta_\star$ &
$\mathcal{U}[0.0,3.0]$ \\
$\log_{10}\epsilon_0$ &
$\mathcal{U}[-3.0,0.0]$ \\
$\log_{10}M_0/M_{\odot}$ &
$\mathcal{U}[7.0,11.0]$ \\
$\sigma_{\rm UV}(10^{10.5}M_\odot)$ &
$\mathcal{U}[0.001,3.0]$ \\
$M$ &
$\mathcal{U}[-21,-18]$ \\
\hline\hline
\end{tabular}

\caption{Summary of the parameter priors used in our analysis.}
\label{tab:prior_summary}

\vspace{0.3cm}

\begin{ruledtabular}
\begin{tabular*}{0.85\textwidth}{@{\extracolsep{\fill}}llcccc}
Interaction
& Parameter
& \multicolumn{2}{c}{\textit{Planck}}
& \multicolumn{2}{c}{\textit{Planck}+JWST} \\
\cline{3-4} \cline{5-6}
&
& Mean & 95\% C.L. upper limit
& Mean & 95\% C.L. upper limit \\
\hline
3I + 0f
& $\mathrm{Log}_{10}(G_{\mathrm{eff}}/\mathrm{MeV}^{-2})$
& -4.61 & -3.29 & -4.81 & -3.41\footnote{Our results for the three-flavor interacting case show only a marginal improvement over the constraints from \textit{Planck} data. This can be attributed to the fact that the 3I + 0f case not only leads to stronger suppression but also produces a larger enhancement. We therefore primarily focus on the one-flavor interacting case in this work.} \\
& $\sum m_\nu\,[\mathrm{eV}]$
& 0.15 & 0.32 & 0.11 & 0.21 \\
\hline
1I + 2f
& $\mathrm{Log}_{10}(G_{\mathrm{eff}}/\mathrm{MeV}^{-2})$
& -4.25 & -2.00 & -4.37 & -2.75 \\
& $\sum m_\nu\,[\mathrm{eV}]$
& 0.17 & 0.46 & 0.12 & 0.24 \\
\end{tabular*}
\end{ruledtabular}

\caption{Summary of the posterior mean values and 95\% C.L. upper limits
obtained from the \textit{Planck}-only and \textit{Planck}+JWST analyses
for the 3I + 0f and 1I + 2f interaction scenarios.}
\label{tab:constraints_summary_combined}

\end{table*}
\clearpage
\onecolumngrid
\begin{center}
\textbf{\large \mytitle}

\vspace{0.08in}
{\it \large Supplemental Material}\\
\vspace{0.12in}

{Ranjini Mondol, Priyank Parashari, Anirban Das, Vera Gluscevic, and Ranjan Laha}
\end{center}

\setcounter{figure}{0}
\setcounter{section}{0}

\makeatletter

\setcounter{equation}{0}
\renewcommand{\theequation}{S\arabic{equation}}
\renewcommand{\thefigure}{S\arabic{figure}}

\renewcommand{\theHfigure}{suppfigure.\arabic{figure}}

\makeatother

\begin{figure}[h!]
    \centering
    \includegraphics[width=0.95\textwidth]{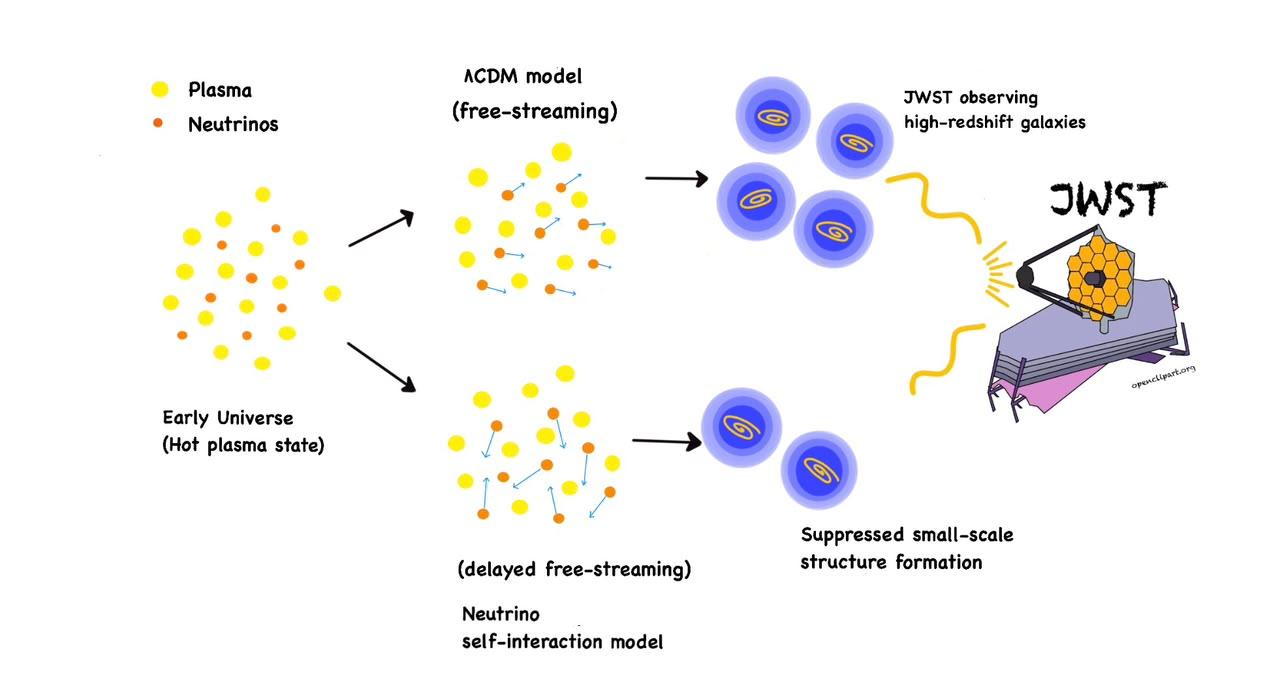}
    \caption{A schematic figure summarizing our results. If neutrinos have self-interactions, they modify structure formation relative to the $\Lambda$CDM model due to delayed free-streaming. JWST UVLF measurements can constrain a range of $G_{\rm eff}$ that suppresses structure formation on the length scales probed by JWST. }
    \label{fig:S1}
\end{figure}

\twocolumngrid

\makeatother
\section{Probing Neutrino self-interactions from UVLF data}

In this work, we use the JWST UVLFs as a probe of BSM neutrino self-interactions ($\nu$SI), \textit{for the first time}. These interactions delay the onset of neutrino free streaming, resulting in a scale-dependent change in the matter power spectrum, $P(k)$, characteristic to the $\nu$SI strength. High-redshift UVLF measurements from JWST probe structure formation on small scales and are therefore sensitive to interaction strengths that modify $P(k)$ on these scales. However, $\nu$SI leaves imprints on $P(k)$ across both large and small scales and changes the temperature and polarization power spectra. We include \textit{Planck} data in our analysis to partially break the degeneracy between astrophysics and cosmology.

\begin{figure}[t]
\begin{flushleft}
\includegraphics[width=1.03\columnwidth]{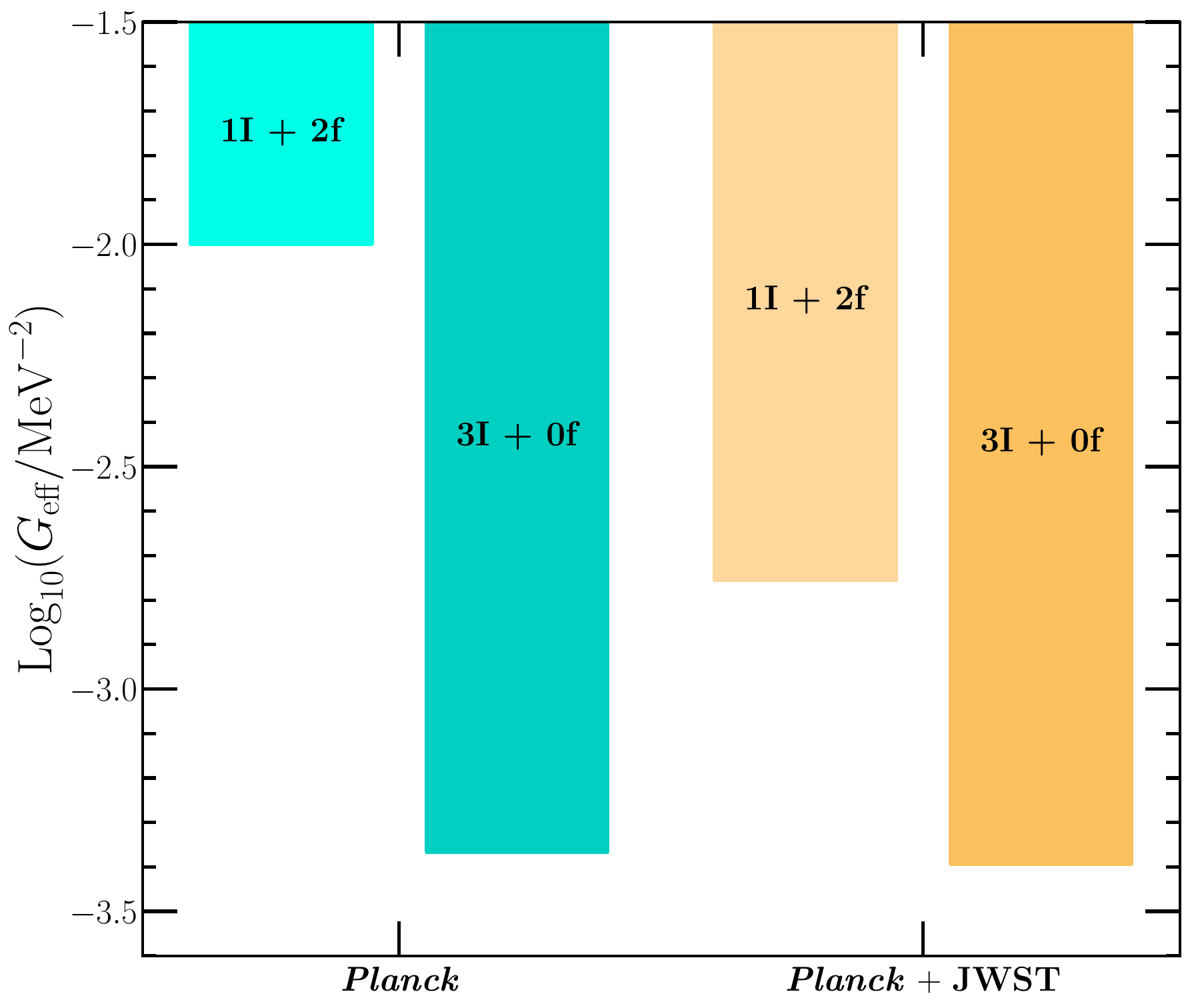}
\end{flushleft}
    \caption{Bar plot comparing the 95\% C.L. upper limits on the BSM neutrino self-interaction parameter, $\mathrm{Log}_{10}(G_{\rm eff}/{\rm MeV}^{-2})$, obtained from the \textit{Planck}-only analysis and from the combined \textit{Planck}+JWST analysis, with the sum of neutrino masses fixed at $\sum m_{\nu}=0.06\,{\rm eV}$. We show results for both the flavor-specific $1{\rm I}+2{\rm f}$ scenario, in which one neutrino flavor is self-interacting, and the flavor-universal $3{\rm I}+0{\rm f}$ scenario, in which all three neutrino flavors are self-interacting. The lighter shaded bars correspond to the $1{\rm I}+2{\rm f}$ case, while the darker shaded bars correspond to the $3{\rm I}+0{\rm f}$ case. The combined \textit{Planck}+JWST analysis gives the 95\% C.L. upper limits $G_{\rm eff}<10^{-2.75}\,{\rm MeV}^{-2}$ for the flavor-specific interaction and $G_{\rm eff}<10^{-3.39}\,{\rm MeV}^{-2}$ for the flavor-universal interaction. The corresponding \textit{Planck}-only limits are
    $G_{\rm eff}<10^{-2.00}\,{\rm MeV}^{-2}$ and $G_{\rm eff}<10^{-3.36}\,{\rm MeV}^{-2}$, respectively.
}
    \label{fig:model-constraint-bar}
\end{figure}

\begin{figure*}
	\begin{center}
		\includegraphics[width=0.98\columnwidth]{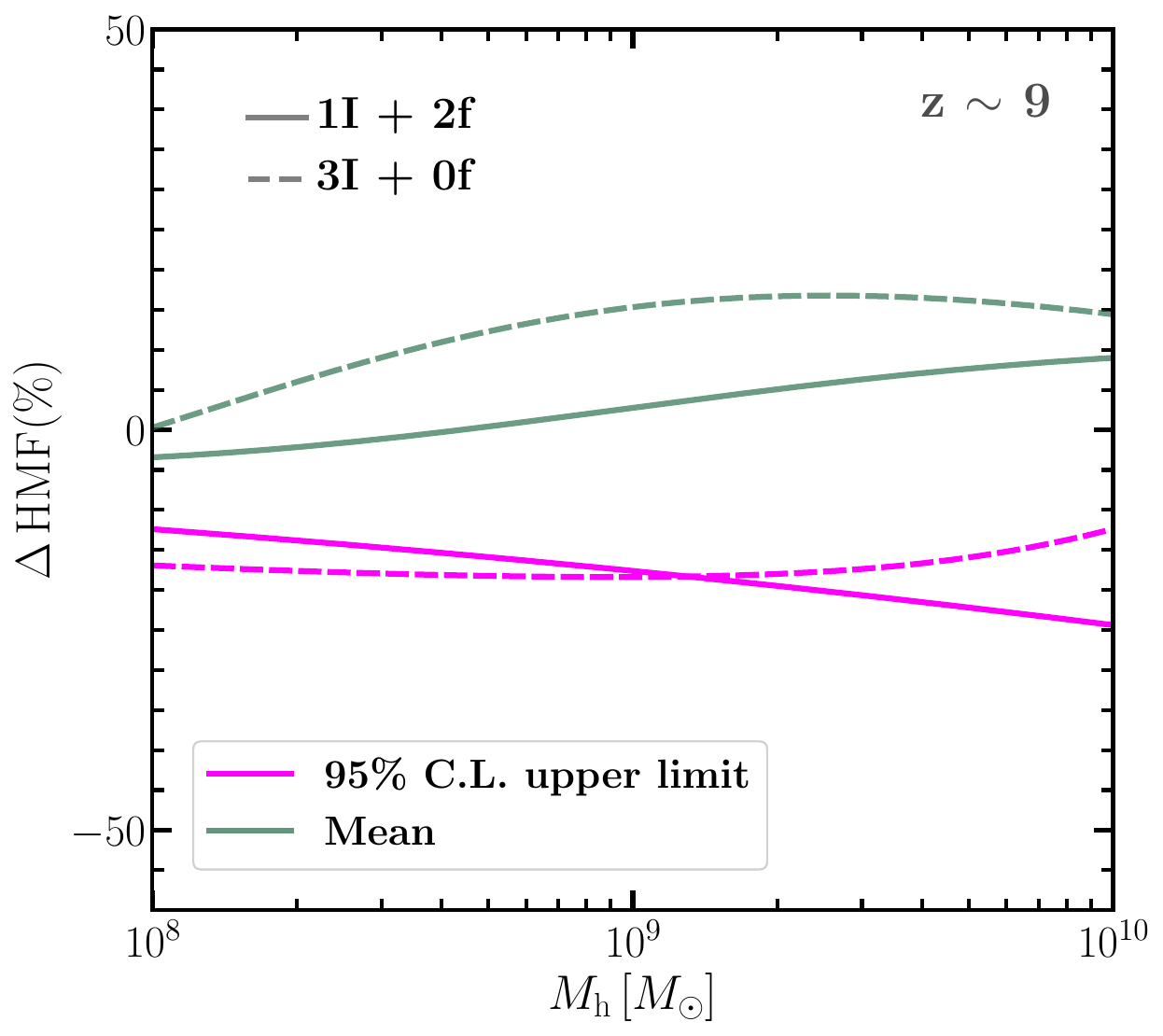}~~
		\includegraphics[width=0.98\columnwidth]{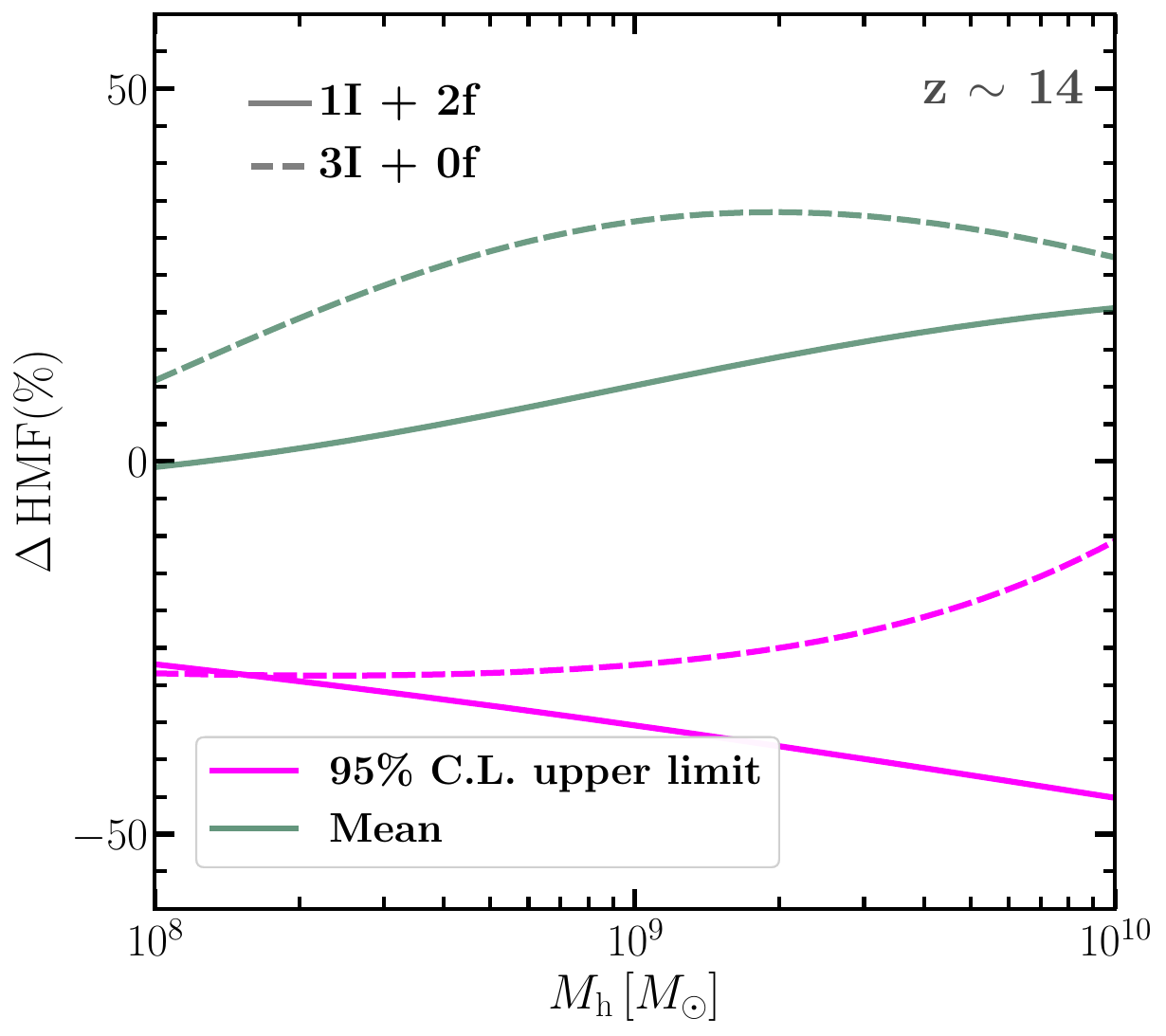}~~\\	
		\caption{Relative change in the halo mass function for $\nu\mathrm{SI} + \sum m_{\nu}$ with respect to the corresponding $\Lambda \mathrm{CDM} + \sum m_{\nu}$ model, $\Delta{\rm HMF}(\%)$. \textit{\textbf{Left Panel:}} $\Delta \mathrm{HMF}(\%)$ evaluated at $z=9$, corresponding to the average redshift of the lowest-redshift spectroscopic UVLF bin used in this work. The solid and dashed lines correspond to the 1I + 2f and 3I + 0f self-coupling cases, respectively. The magenta and green curves are evaluated using the 95\% C.L. upper limit values and the mean values, respectively, derived from the MCMC posteriors for $G_{\mathrm{eff}}$ and $\sum m_\nu$.
        For the 95\% C.L. upper limit curves, we use $\mathrm{Log}_{10}(G_{\mathrm{eff}}/{\rm MeV}^{-2})=-2.75$ and $\sum m_\nu=0.24\,{\rm eV}$ for 1I + 2f, and $\mathrm{Log}_{10}(G_{\mathrm{eff}}/{\rm MeV}^{-2})=-3.41$ and $\sum m_\nu=0.21,{\rm eV}$ for 3I + 0f. \textit{\textbf{Right Panel:}} Same as the left panel, but evaluated at $z=14$, corresponding to the highest average redshift in our spectroscopic dataset.
For both panels, we fix the six $\Lambda$CDM parameters to the values from the TT, TE, EE+lowE+lensing analysis~\cite{Planck:2018vyg}.
    }
		\label{fig:HMF_two_redshift}
	\end{center}	
\end{figure*}

\begin{figure*}
	\begin{center}
		\includegraphics[width=0.98\columnwidth]{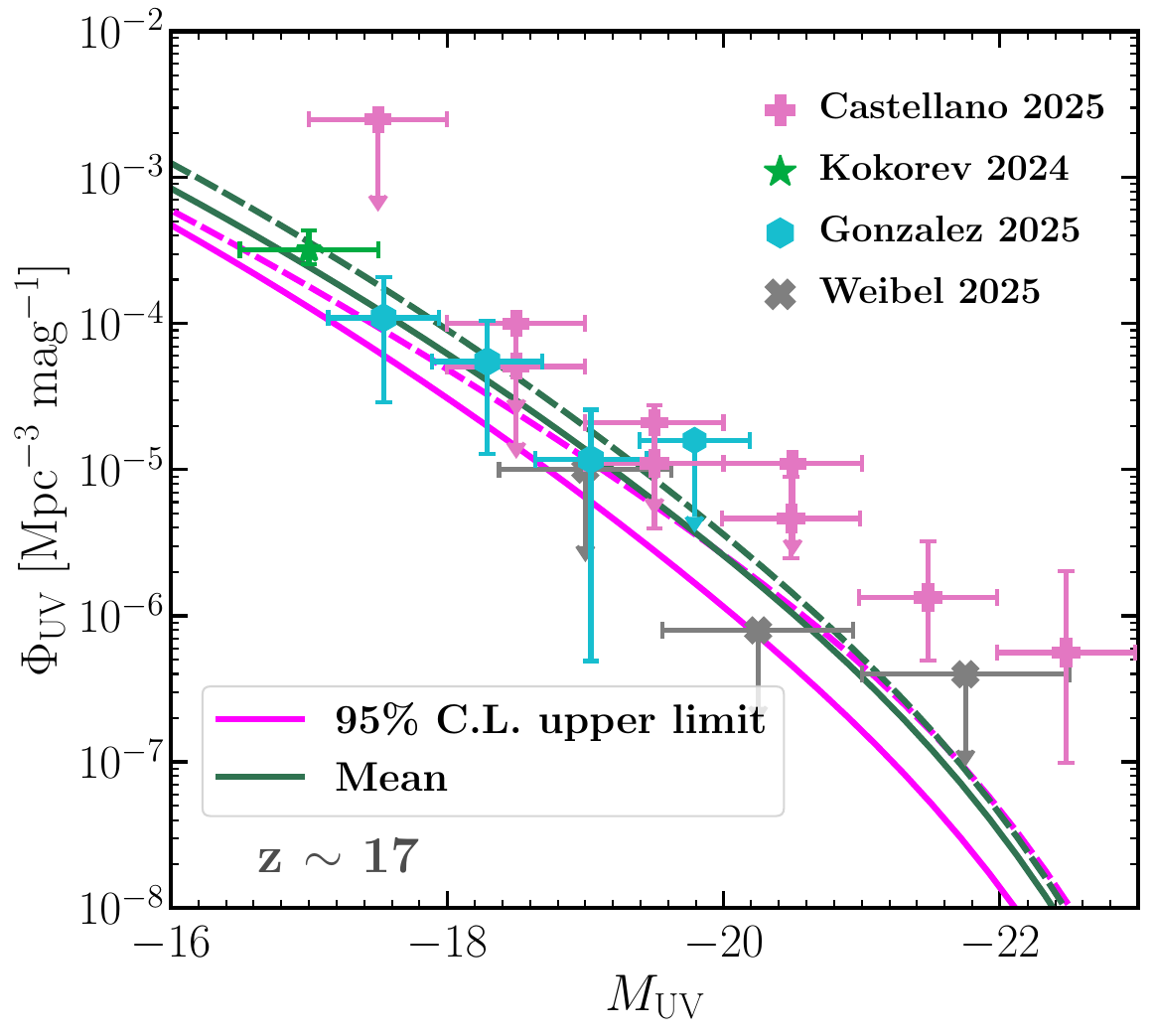}~~
		\includegraphics[width=1.02\columnwidth]{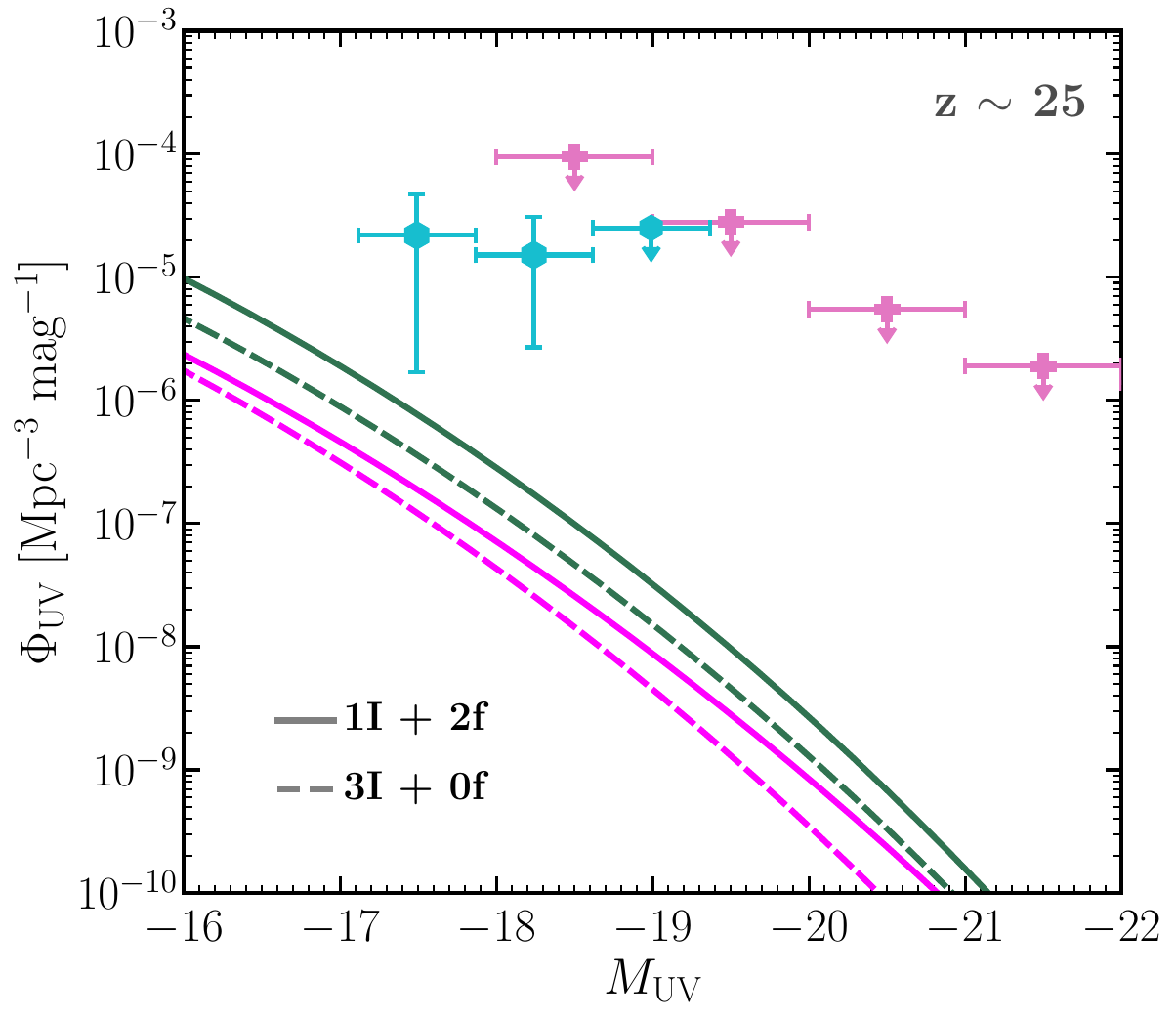}~~\\	
		\caption{\textit{\textbf{Left Panel:}} UV luminosity function evaluated at $z\sim17$. 
The observational data shown are photometric measurements listed in the \textit{End Matter}. 
The solid and dashed curves correspond to the UVLFs calculated for the 1I + 2f and 3I + 0f self-coupling cases, respectively. 
The magenta and green curves are evaluated using the 95\% C.L. upper limit values and the mean values, respectively, derived from the MCMC posteriors for $G_{\mathrm{eff}}$ and $\sum m_\nu$. 
For the 95\% C.L. upper limit curves, we use $\mathrm{Log}_{10}(G_{\mathrm{eff}}/{\rm MeV}^{-2})=-2.75$ and $\sum m_\nu=0.24\,{\rm eV}$ for 1I + 2f, and $\mathrm{Log}_{10}(G_{\mathrm{eff}}/{\rm MeV}^{-2})=-3.41$ and $\sum m_\nu=0.21\,{\rm eV}$ for 3I + 0f.
\textit{\textbf{Right Panel:}} Same as the left panel, but evaluated at $z\sim25$. 
For both panels, we fix the six $\Lambda$CDM parameters to the values from the TT, TE, EE+lowE+lensing analysis~\cite{Planck:2018vyg}. 
To isolate the effects of $G_{\mathrm{eff}}$ and $\sum m_\nu$ on the UV luminosity function, we fix the astrophysical parameters for all curves from at their mean values from the Planck+JWST posterior for the flavor-universal 3I + 0f case. The values of the astrophysical parameters are 
$\{\alpha_\star,\beta_\star,\mathrm{Log}_{10}\epsilon_0,\mathrm{Log}_{10}(M_0/M_\odot),\sigma_{\rm UV, 10.5}\}
= \{0.81,0.71,-2.31,8.87, 0.83\}$, 
where $\alpha_\star$, $\beta_\star$, $M_0$, $\sigma_{\mathrm{UV}}$, and $\epsilon_0$ are the SFE model parameters.
    }
		\label{fig:uvlf}
	\end{center}	
\end{figure*}

In this \textit{Supplemental Material}, we present our analysis in greater detail. In Fig.~\ref{fig:model-constraint-bar}, we schematically show the constraints obtained from MCMC runs with fixed neutrino mass. Constraints from the \textit{Planck}-only analysis are shown as green-shaded bars, whereas those from the combined \textit{Planck}+JWST analysis are shown as yellow-shaded bars. For both the flavor-universal and flavor-specific cases, including the JWST likelihood strengthens the constraints. Figure~\ref{fig:HMF_two_redshift} shows $\Delta\mathrm{HMF}(\%)$ arising from the combined effects of $G_{\mathrm{eff}}$ and $\sum m_{\nu}$ at $z=9$ and $z=14$, the lowest and highest average redshifts in our spectroscopic dataset. In Fig.~\ref{fig:uvlf}, we show the effects of $G_{\rm eff}$ and $\sum m_{\nu}$ on the UVLFs evaluated at the highest median redshifts considered in this work, while Fig.~\ref{fig:planck_tt_ee_combined} shows their effects on the \textit{Planck} TT and EE power spectra. Fig.~\ref{fig:neutrino-mass-posterior} presents the 1D posteriors of $\mathrm{Log}_{10}(G_{\mathrm{eff}}/\mathrm{MeV}^{-2})$ and $\sum m_{\nu}$ for the flavor-universal case, comparing the \textit{Planck}-only and \textit{Planck}+JWST analyses. For the flavor-specific case, the posterior of $\sum m_{\nu}$ is shown in Fig.~\ref{fig:sum_mnu_1_flavor}. Figs.~\ref{fig:2D_posterior3I0f} and~\ref{fig:2D_posterior1I2f} present the 1D and 2D marginalized posteriors of the astrophysical parameters from the \textit{Planck}+JWST analyses for flavor universal and flavor specific cases, respectively. Figs.~\ref{fig:2D_posterior3I0f-cosmo} and \ref{fig:2D_posterior1I2f-cosmo} show the 1D and 2D marginalized posteriors from different analyses for the three-flavor and one-flavor interaction scenarios, respectively. In Fig.~\ref{fig:model-constraint}, we project our limit on one-flavor interacting case onto the parameter space of a UV-complete model and compare it with existing constraints from the literature.  

UVLFs become sensitive to the $P(k)$ through the halo mass function. The enhancement and suppression induced in $P(k)$ from $\nu$SI are at most 15\%. Since JWST is still in the early stages of observations, and galaxy number densities at high-redshifts are low, the associated uncertainties remain significant.\,\,A larger volume of data can not only reduce the error bars but also can provide us access to higher $k$ values and improve our understanding of high-redshift astrophysics. Finally, in Fig.~\ref{fig:fs_redshift_both}, we discuss the tentative range of $G_{\mathrm{eff}}$ that could be probed with additional high-redshift UVLF measurements in the future.

\section{The Halo Mass Function Calculation}
In this section, we describe the formalism for calculating the mass distribution of halos as a function of redshift, namely the halo mass function (HMF). We have followed the same prescription asin~\cite{Das:2025bnr, Parashari:2023cui}. The HMF is given by~\cite{Das:2025bnr}
\begin{equation}
\frac{dn}{d\, \mathrm{Log}_{10}{M_{h}}} = M_{h}\frac{\rho_{m,0}}{M_{h}^{2}}f(\sigma)\Big|\frac{d \mathrm{Log}{\sigma}}{d\, \mathrm{Log}_{10}{M_{h}}}\Big|\,,
\end{equation}
where $\rho_{m,0}$ is the present-day mean matter density, $M_{h}$ is mass of the halo, and $\sigma^{2}(M_{h})$ is the mass variance~\cite{Das:2025bnr} as also defined in the main text. The HMF, and consequently the UVLF is sensitive to the matter power spectrum through $\sigma^{2}(M_{h})$. In addition, we use the smooth-$k$ window function, which is generalized to portray a wide variety of DM models including CDM~\cite{Leo:2018odn, Verwohlt:2024efh}. We choose the parameters given in~\cite{Verwohlt:2024efh}. Unlike warm dark matter (WDM), interacting dark matter (IDM), and fuzzy dark matter (FDM) models, the matter power spectrum in the neutrino self-interaction model does not exhibit a sharp cutoff but instead shows a $\sim$ 15\% change. Hence, using the filter function from~\cite{Verwohlt:2024efh} seems to be a reasonable choice for our purposes. However, given the current error bars on the JWST data and after marginalization over the astrophysical parameters, we find that our results are largely insensitive to the choice of filter function.

The other semi-analytical function, the fitting function, is taken to be the Sheth-Tormen (ST) fitting function
\begin{equation}
f(\sigma) = A\sqrt{\frac{2a}{\pi}}\Big\{1+ \Big(\frac{\sigma^{2}}{a \delta_{c}^{2}}\Big)^{p}\Big\}\frac{\delta_{c}}{\sigma}\exp{-\frac{a\delta_{c}^{2}}{2\sigma^{2}}}\,.
\end{equation}
The best-fit values of its fitting parameters are $A = 0.3222$, $a = 0.707$, $p = 0.3$, and $\delta_{c} = 1.686$~\cite{Sheth:1999su, Parashari:2023cui}. Several other fitting functions have been calibrated using high-redshift simulations~\cite{Murray:2013qza}. However, since the ST fitting function is widely used in the $\Lambda$CDM case and is analytically motivated, we use it in this work. A more appropriate approach would be to use results from simulations that explicitly include such interactions. 

In Fig.~\ref{fig:HMF_two_redshift}, we plot the percentage change in the HMF due to $\nu$SI, defined as $\Delta \mathrm{HMF}(\%) = 100 \times (\mathrm{HMF}_{\nu\mathrm{SI}} - \mathrm{HMF}_{\Lambda\mathrm{CDM}, m_{\nu}})/\mathrm{HMF}_{\Lambda\mathrm{CDM},m_{\nu}}$, at the median redshifts of the lowest and highest redshift bins for the spectroscopically confirmed galaxies, $z \sim 9$ (left panel) and  $z \sim 14$ (right panel), respectively. The solid (dashed) lines correspond to $\Delta \mathrm{HMF} (\%)$ for flavor-specific (flavor-universal) couplings, evaluated at the mean and 95\% C.L. upper limits of the parameters $G_{\mathrm{eff}}$ and $\sum m_{\nu}$, shown in green and magenta, respectively. These results show that, for both 3I + 0f and 1I + 2f, the mean values lead to an enhancement, whereas the curves corresponding to the 95\% excluded values show suppression in both cases. This pattern is consistent with the high-redshift behaviour shown in the right panel of Fig.~2 in the main text. At lower redshifts, this effect is comparatively smaller, suggesting that the high-redshift data play an important role in driving the constraints.

\section{The Galaxy-Halo Connection Prescription}

We see the luminous matter, not the DM halo that provides the scaffolding for these luminous galaxies. In this section, we connect the mass of a DM halo, $M_{h}$, to the brightness of the galaxy hosted by a DM halo, namely UV absolute magnitude, $M_{\mathrm{UV}}$. We use the following formalism to calculate the SFR and, eventually, the UV luminosity function. We adapt the same formalism also described in~\cite{Das:2025bnr}.

Due to the gravitational potential sourced by a DM halo, baryons accrete onto it and eventually form stars. Assuming adiabatic initial conditions, the baryonic accretion rate is related to the halo mass accretion rate through the cosmic baryon fraction $f_{b} = \Omega_{b}/\Omega_{m} \approx 0.16$. Here $\Omega_{m}$ and $\Omega_{b}$ are the dimensionless density parameters of the total pressureless matter and the baryons, respectively.\,\,The star formation rate is then written as~\cite{Sabti:2021xvh, Das:2025bnr, Shen:2024hpx}
\begin{equation}
\text{SFR}=\epsilon_\star f_b \dot{M}_h ,\label{eq}
\end{equation}

where $\epsilon_{\star}$ is the star formation efficiency and $\dot{M}_{h}$ is the halo mass accretion rate. The halo-mass-dependent star formation efficiency used in our work follows a framework adopted in~\cite{Das:2025bnr, Shen:2025isu} and is given by

\begin{equation}
\epsilon_\star(M_h)= \frac{\epsilon_{0}}{(M_h/M_{0})^{-\alpha_\star} + (M_h/M_{0})^{-\beta_\star}},\label{eq:SFE}
\end{equation}
where $\epsilon_{0}$, $M_{0}$, $\alpha_{\star}$, and $\beta_{\star}$ are model parameters that are marginalized over to obtain our results. This particular formalism, presented in~\cite{Shen:2025isu}, is motivated to phenomenologically capture the effects of gas dynamics at very high redshifts ($z\gtrsim 10$).

The parametric form of the $\dot{M}_{h}$ is obtained from \textsc{GUREFT} N-body simulations~\cite{Yung:2023sib}

\begin{eqnarray}
\frac{\dot{M}_h}{\rm{M_\odot~yr^{-1}}} &= \beta(z)(M_{h, 12}E(z))^{\alpha(z)} \nonumber \,\,,\\
\alpha(z) &= 0.858 + 1.554 a - 1.176 a^{2}\,\,, \nonumber\\
\mathrm{Log}\,{\beta(z)}  &= 2.578-0.989 a -1.545 a^{2},   \nonumber
\label{eq:HaloAccRate}
\end{eqnarray} 
where $E(z) = H(z)/H_{0}$, $M_{h, 12} = M_{h}/10^{12}M_{\odot}$, and $a = 1/(1+z)$, the scale factor. The UV luminosity of the galaxies has an empirical relation with the SFR, calibrated with the simulations of stellar populations~\cite{Sabti:2021xvh}

\begin{figure*}
\centering
\begin{minipage}{0.48\textwidth}
    \centering
    \includegraphics[width=\linewidth]{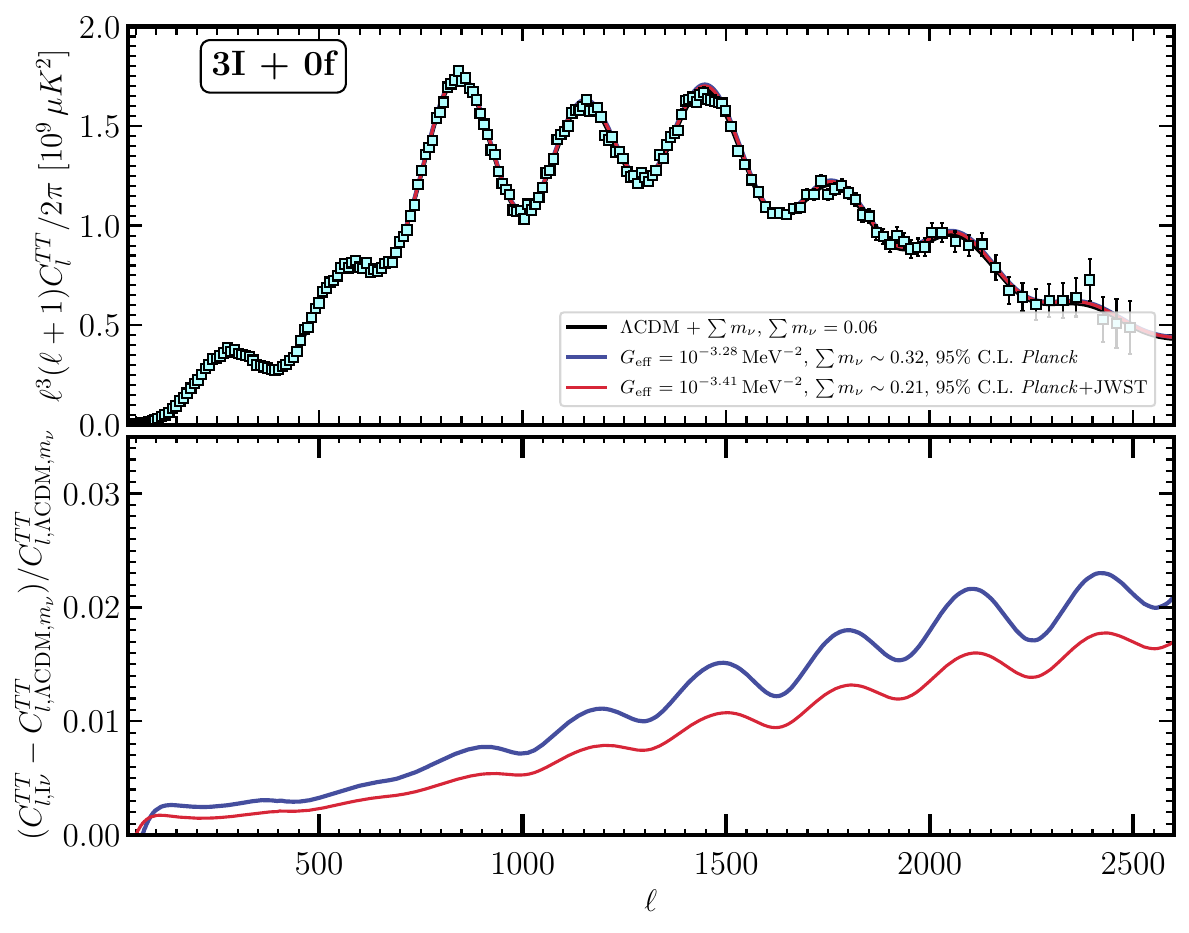}
\end{minipage}
\hfill
\begin{minipage}{0.48\textwidth}
    \centering
    \includegraphics[width=\linewidth]{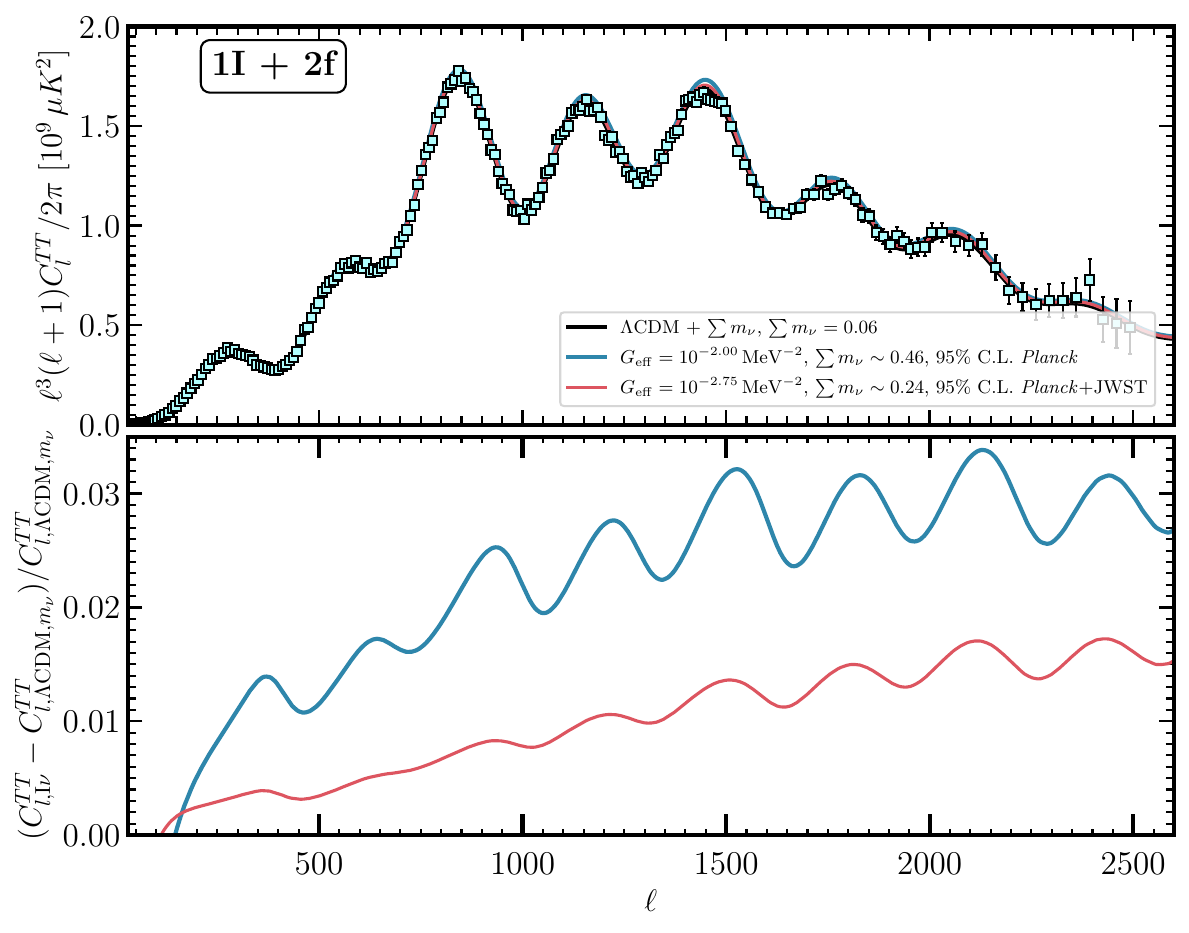}
\end{minipage}

\vspace{0.4cm}

\begin{minipage}{0.48\textwidth}
    \centering
    \includegraphics[width=\linewidth]{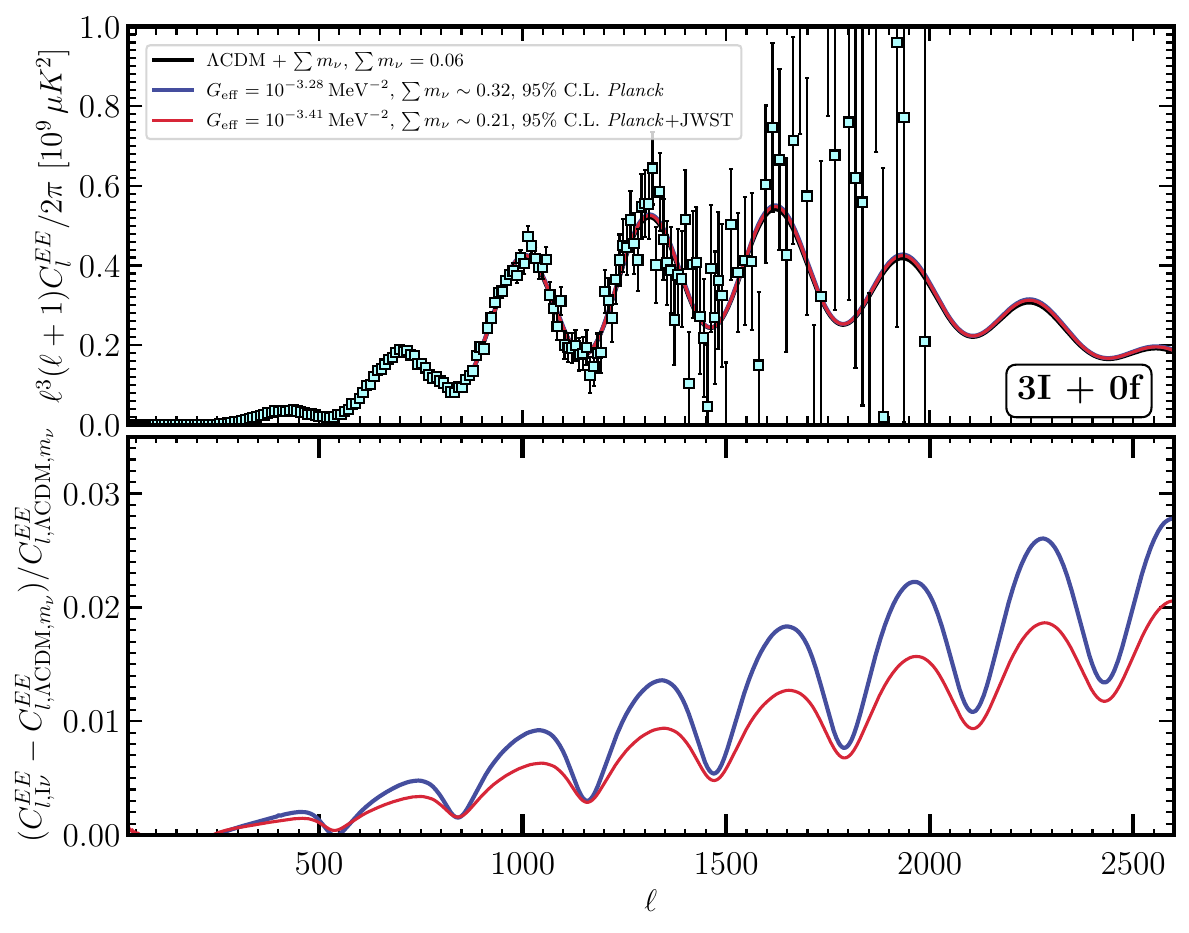}
\end{minipage}
\hfill
\begin{minipage}{0.48\textwidth}
    \centering
    \includegraphics[width=\linewidth]{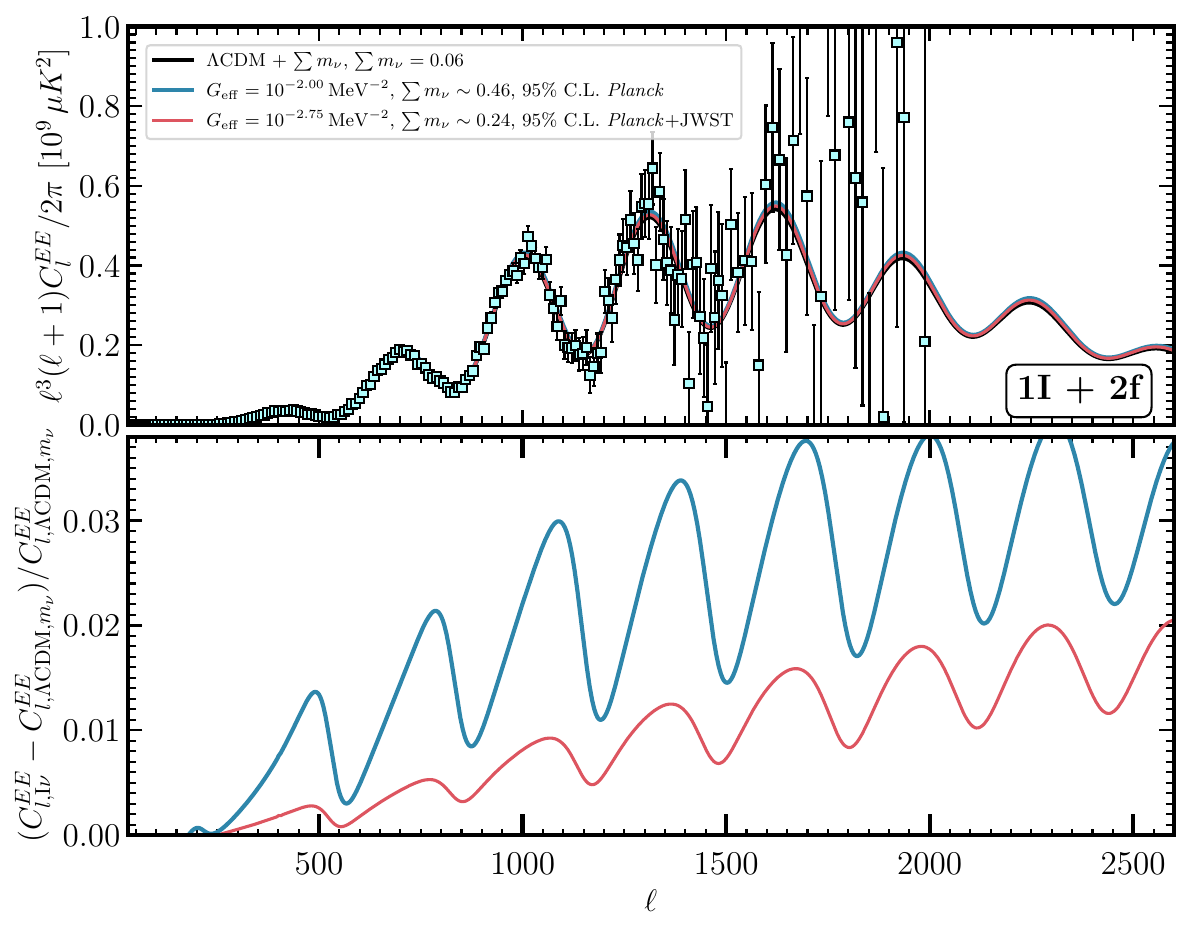}
\end{minipage}

\caption{Effects of $G_{\mathrm{eff}}$ and $\sum m_\nu$ on the amplitude and phase of the CMB temperature and polarization spectra. 
In each plot, the upper panel shows $\ell^3(\ell+1)C_\ell^{XX}/(2\pi)$, while the lower panel shows the fractional change relative to the fiducial $\Lambda$CDM+$\sum m_\nu$ model, $(C_{\ell}^{XX}-C_{\ell,\Lambda{\rm CDM}}^{XX})/C_{\ell,\Lambda{\rm CDM}}^{XX}$, where $X=\{T,E\}$. 
The left panels correspond to the 3I + 0f self-interaction case, whereas the right panels correspond to the 1I + 2f case. 
In each panel, the blue and red curves are evaluated using the 95\% C.L. upper limit values of $G_{\mathrm{eff}}$ and $\sum m_\nu$ obtained from the \textit{Planck}-only analysis and the combined \textit{Planck}+JWST analysis, respectively. 
The remaining six $\Lambda$CDM parameters are fixed to their best-fit values from the TT+TE+EE+BAO+$H_0$ dataset. 
The fiducial $\Lambda$CDM+$\sum m_\nu$ curves are evaluated at $\sum m_\nu=0.06\,{\rm eV}$.}
\label{fig:planck_tt_ee_combined}

\end{figure*}

\begin{equation}
    L_{\nu}^{\mathrm{UV}} = \frac{1}{\kappa_{\mathrm{UV}}}\,\mathrm{SFR} \,.
\end{equation}

The value of the constant of proportionality $\kappa_{\mathrm{UV}}$ depends on the stellar initial mass function. Here, we use $\kappa_{\mathrm{UV}} = 0.72 \times 10^{-28}\,\mathrm{M}_{\odot}\mathrm{yr}^{-1}\mathrm{erg}^{-1}s$ from~\cite{Shen:2024hpx}. The UV magnitude is related to the luminosity as
\begin{equation}
    \mathrm{Log}_{10}\Big(\frac{L_{\nu}^{\mathrm{UV}}}{\mathrm{erg\,s^{-1}}}\Big) = 0.4(51.63-M_{\mathrm{UV}}) \,.\label{eq:lum-eq}
\end{equation}

This magnitude $M_{\mathrm{UV}}$ obtained from Eq.~\ref{eq:lum-eq} is the median UV-magnitude for the galaxy hosted by a DM halo of mass $M_{h}$. However, the $M_{\mathrm{UV}}-M_{h}$ relation is not completely deterministic and shows a scatter because star formation is a stochastic process. This scatter we model in the same way as~\cite{Das:2025bnr}

\begin{equation}
    p(M_{\rm UV}|M_h) = \frac{1}{\sqrt{2\pi}\sigma_{\rm UV}}
    \exp\left[-\frac{\left(M_{\rm UV}-M_{\rm UV}^{\rm med}\right)^2}{2\sigma_{\rm UV}^2}\right]\,,
\end{equation}
where $p(M_{\mathrm{UV}}|M_{h})$ is the probability distribution of magnitude of galaxies $M_{\mathrm{UV}}$ emitted from a DM halo of mass $M_{h}$. We use a Gaussian kernel centered around a median magnitude $M_{\mathrm{UV}}^{\mathrm{med}}$ to model this probability distribution. The width $\sigma_{\rm UV}(M_h)$, termed UV variability
\begin{eqnarray}
\sigma_{\rm UV}(M_h) = \text{max}[\sigma_{\rm UV, 10.5} - 0.34 \mathrm{Log}_{10}(M_{h, 10.5}), 
0.4]\,, \phantom{xxx}
\end{eqnarray}
where $\sigma_{\rm{UV, 10.5}} $ is the UV variability at a DM halo mass of $10^{10.5}M_{\odot}$ and $M_{h, 10.5}$ is the halo mass scaled by $10^{10.5}M_{\odot}$. Our astrophysical model parameters are assumed to be independent of the redshift, so the entire $z$-dependence of the UVLF arises from the HMF.

At high redshifts, the impact of dust attenuation on the UVLF becomes negligible, since the astrophysical processes that contribute to dust are highly suppressed at such early epochs. We take the effect of dust attenuation into account in the same way as in~\cite{Das:2025bnr}.

In Fig.~\ref{fig:uvlf}, we show the observable UVLFs computed at the mean values (green curves) and at the 95\% C.L.\ upper limits (magenta curves) of $G_{\mathrm{eff}}$ and $\sum m_{\nu}$, for the two highest average redshifts used in this work. All the data points shown here are photometric. Evaluating the UVLF also requires astrophysical inputs, and given the current error bars, it is difficult to clearly distinguish between the mean and statistically rejected values. In these plots, we use the same astrophysical parameters to compute the theoretical binned UVLFs, namely the mean values obtained from the \textit{Planck}+JWST flavor-universal MCMC run using the {\tt GetDist} analysis. This choice is reasonable because the mean astrophysical parameters agree with each other within the $1\sigma$ range. As expected from the behavior of the HMF, the UVLFs computed using the 95\% C.L.\ rejected upper limits are suppressed compared to those evaluated using the mean values.

We have only one Universe to observe, and even at low redshifts—where a wealth of data is available—we still face significant challenges in fully understanding the complex dynamics of baryons. With JWST, we are now directly probing these early epochs in the history of our Universe, where our understanding of the underlying astrophysical processes remains limited. As a result, we must rely on a limited set of simulations specifically tailored to study these epochs. In principle, tighter constraints could be obtained by including UVLF data from HST~\cite{Sabti:2021xvh}, which have much smaller error bars. However, the relevant astrophysical processes evolve with redshift, and this redshift dependence must be incorporated into the astrophysical models~\cite{Blamart:2025szc,2026arXiv260530436H,Samuel:2026ymj}. In addition, modeling the effects of dust at lower redshifts introduces further challenges~\cite{Shen:2021crm,Sabti:2021xvh}. We account for these astrophysical uncertainties by marginalizing over the astrophysical parameters and derive our constraints from the resulting posteriors.

\begin{figure*}
	\begin{center}
		\includegraphics[width=\columnwidth]{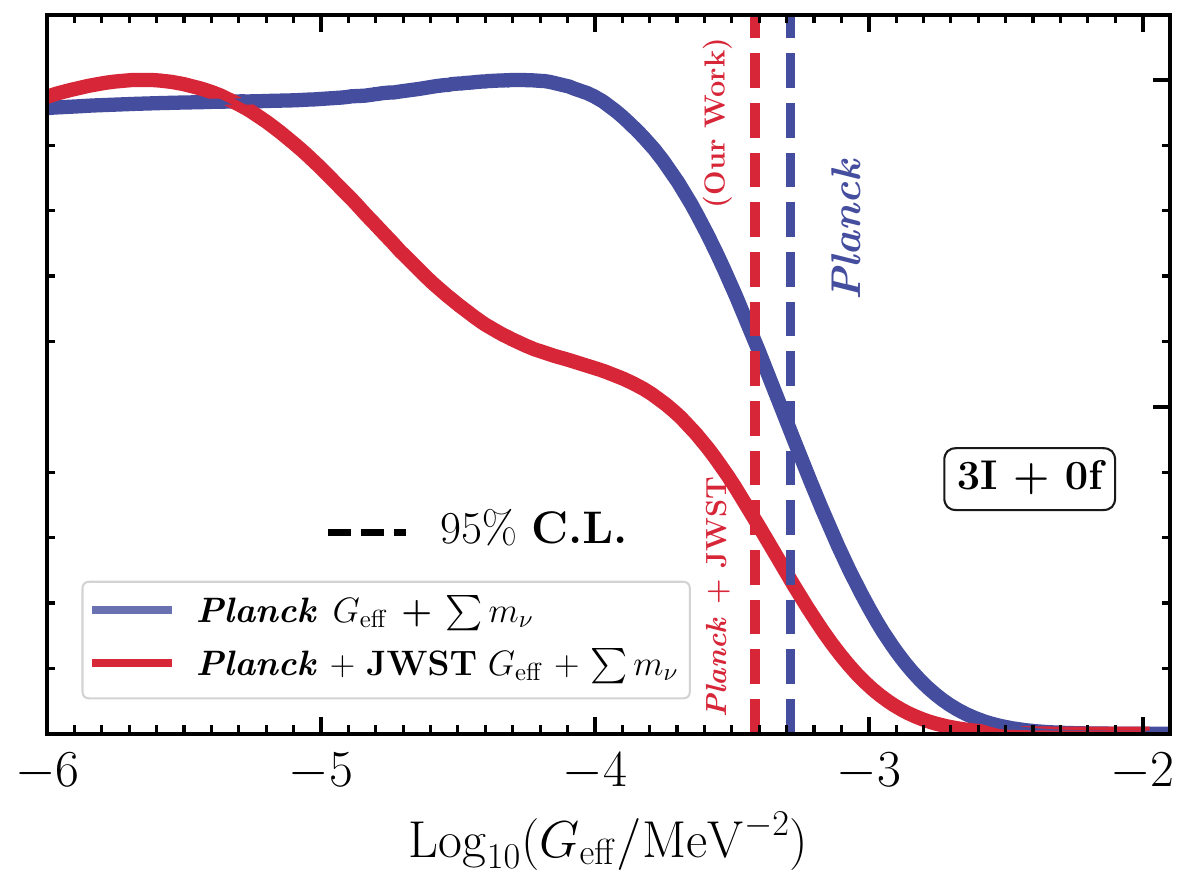}
		\includegraphics[width=\columnwidth]{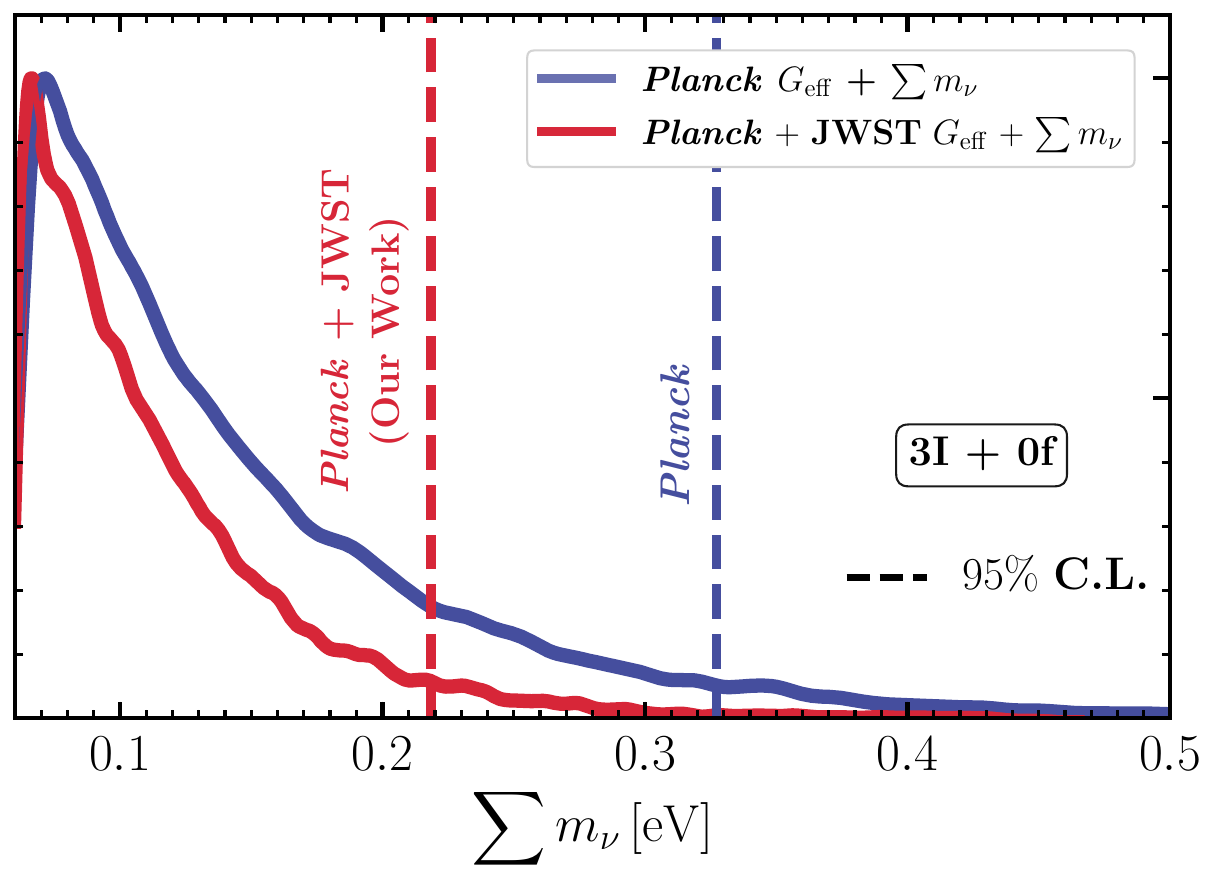}~~\\	
		\caption{\textit{\textbf{Left Panel:}} Comparison of the marginalized posterior distributions of the interaction strength 
$\mathrm{Log}_{10}(G_{\mathrm{eff}}/{\rm MeV}^{-2})$ for the 3I + 0f coupling scenario, obtained from the \textit{Planck}-only analysis 
(blue solid line) and from the combined \textit{Planck}+JWST analysis (red solid line). 
The dashed vertical lines show the corresponding 95\% C.L. upper limits, 
$\mathrm{Log}_{10}(G_{\mathrm{eff}}/{\rm MeV}^{-2}) < -3.29$ for \textit{Planck}-only and 
$\mathrm{Log}_{10}(G_{\mathrm{eff}}/{\rm MeV}^{-2}) < -3.41$ for \textit{Planck}+JWST. 
\textit{\textbf{Right Panel:}} Comparison of the marginalized posterior distributions of $\sum m_\nu$ for the same two datasets, using the same color coding. 
The lower bound of the prior on $\sum m_\nu$ is set to $0.06\,{\rm eV}$. 
The 95\% C.L. upper limit on $\sum m_\nu$ is $0.32\,{\rm eV}$ for \textit{Planck} only and $0.21\,{\rm eV}$ for \textit{Planck}+JWST. 
Thus, adding JWST data to \textit{Planck} leads to a marginally tighter constraint on 
$\mathrm{Log}_{10}(G_{\mathrm{eff}}/{\rm MeV}^{-2})$ and a substantial improvement in the constraint on $\sum m_\nu$.}
		\label{fig:neutrino-mass-posterior}
	\end{center}	
\end{figure*}

\section{Effect on the cosmological observables}

The delay in neutrino free streaming due to $\nu$SI affects the evolution of photon-baryon perturbation modes that enter the horizon before neutrinos begin to free-stream. Within the $\Lambda$CDM, neutrinos begin to free-stream at a temperature of $\sim 1\,\mathrm{MeV}$ \cite{Kreisch:2022zxp}. In the absence of $\nu$SI, free streaming neutrinos develop anisotropic stress, which gravitationally imprints their signature on the CMB power spectrum, making CMB observables sensitive to such interactions~\cite{Kreisch:2019yzn, Kreisch:2022zxp}.

\begin{figure}[t]
\begin{flushleft}
\includegraphics[width=1.03\columnwidth]{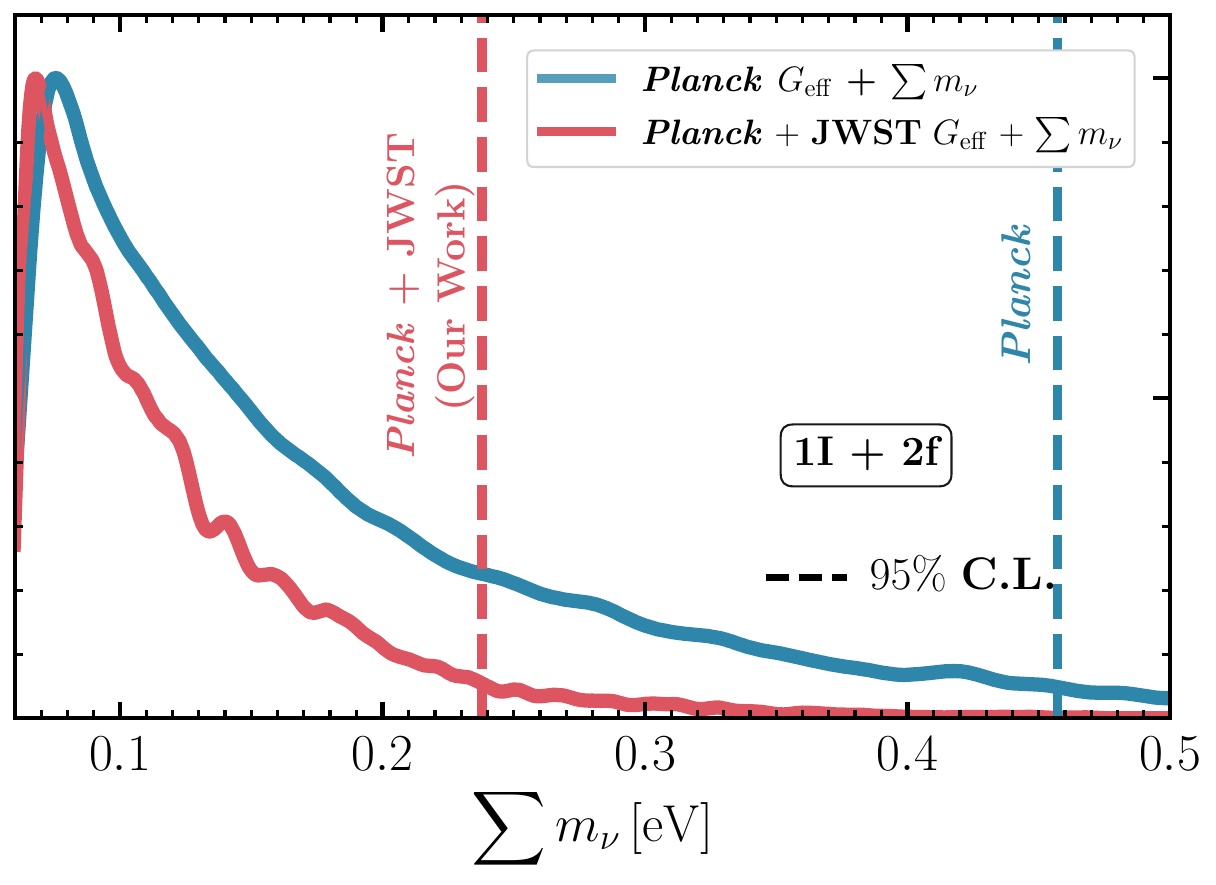}
\end{flushleft}
    \caption{Comparison of the posterior distributions for the $\sum m_{\nu}$ for the $1\mathrm{I}+2\mathrm{f}$ coupling scenario, obtained from the \textit{Planck}-only analysis (blue solid line) and from our analysis using \textit{Planck}+JWST data 
        (red solid line). The dashed vertical lines show the corresponding 95\% C.L. upper limits on the $\sum m_{\nu}$.
}
    \label{fig:sum_mnu_1_flavor}
\end{figure}

\begin{figure*}[t]
    \centering
    \includegraphics[width=\textwidth]{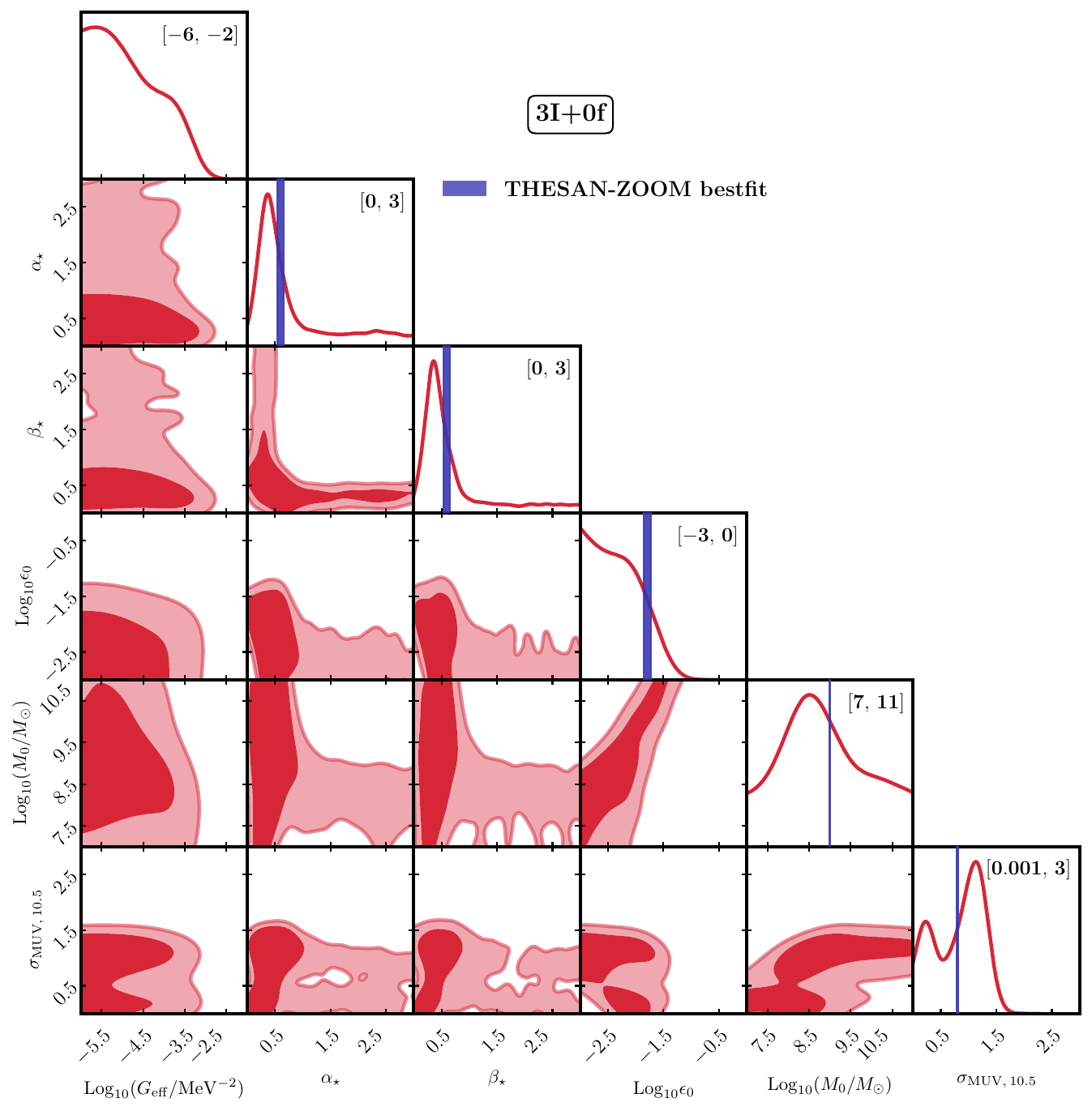}
    \caption{Contours of the 68\% and 95\% credible regions for the $\mathrm{Log}_{10}(\rm GeV/MeV^{-2})$ \& the astrophysical parameters with $\textit{Planck}$+JWST analysis. The diagonal panels show the 1D marginalized posteriors, 
    and the lower off-diagonal panels show the 2D marginalized posteriors. The dark purple shaded region spans the best-fit parameter values over the redshift range $z\sim8\text{--}10$ from Table~3 of Ref.~\cite{Shen:2025isu}. Since Ref.~\cite{Shen:2025isu} adopts a fixed normalization of the variability, $\sigma_{\rm MUV,10.5}=0.8$, consistent with various simulations, we indicate this value with a line. This plot is evaluated for the flavor-universal case.}
    \label{fig:2D_posterior3I0f}
\end{figure*}

Free streaming neutrinos move supersonically through the photon-baryon plasma. Owing to their large energy density, they exert a gravitational pull on the photon-baryon wavefront, pulling it ahead relative to a scenario without free streaming neutrinos. This effect manifests itself as a shift of the CMB power spectra towards smaller $\ell$-modes \cite{Kreisch:2022zxp, Kreisch:2019yzn, Das:2023npl}. Modes entering the horizon before the onset of free streaming do not show this characteristic feature at horizon entry, since neutrinos remain tightly coupled. However, modes that enter the horizon between the onset of free streaming and matter-radiation equality start to show the usual phase shift and amplitude suppression. $\nu$SI affects both the TT and EE power spectra in a qualitatively similar and discussed thoroughly in Ref.~\cite{Kreisch:2019yzn}.

\begin{figure*}[t]
    \centering
    \includegraphics[width=\textwidth]{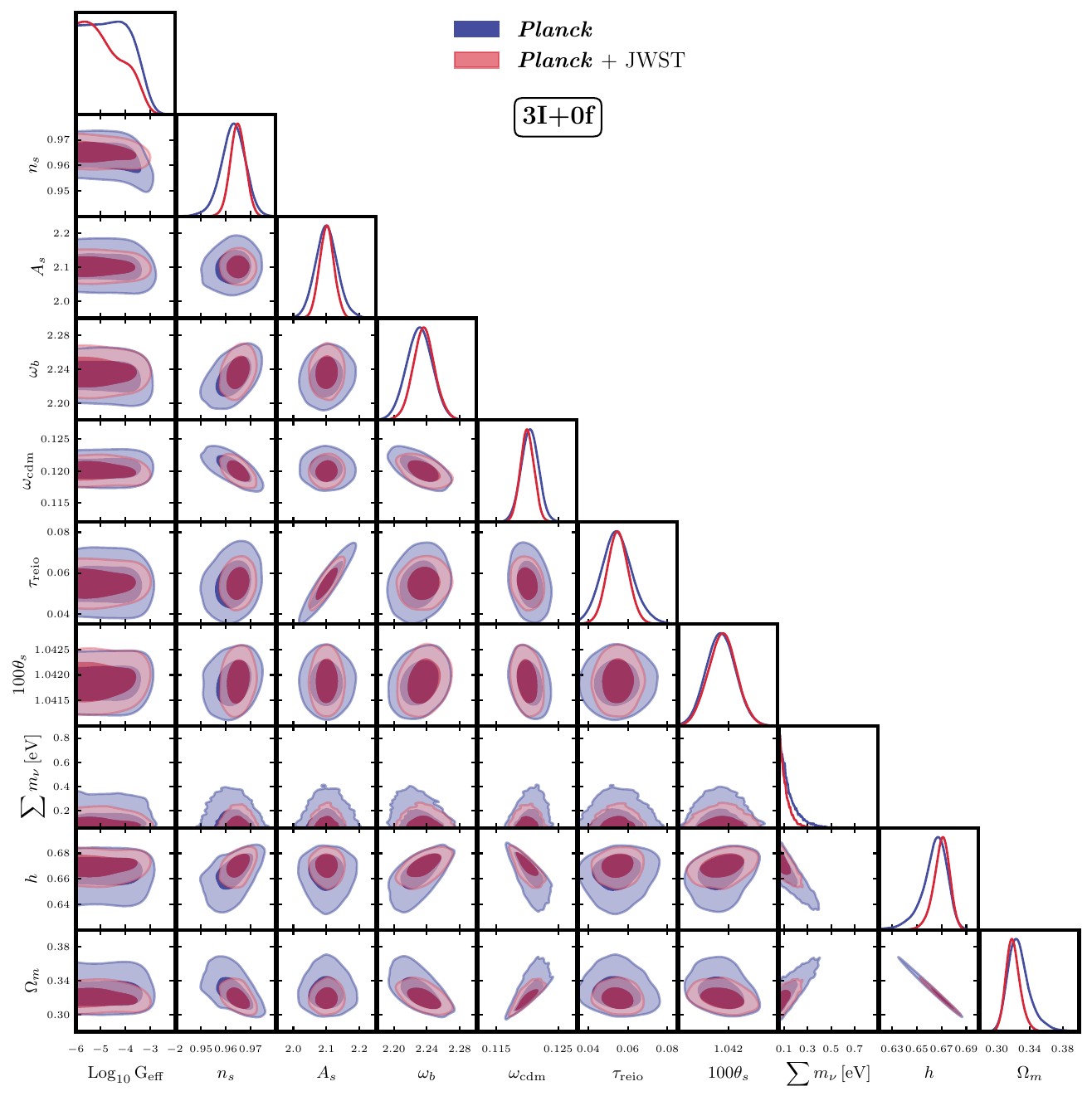}
    \caption{Comparison of the 68\% and 95\% C.L. credible regions for the cosmological parameters in the $G_{\mathrm{eff}}+\sum m_{\nu}$ model. The blue shaded contours correspond to the \textit{Planck}-only analysis, while the red shaded contours correspond to the combined \textit{Planck}+JWST analysis. The diagonal panels show the corresponding 1D marginalized posteriors, 
    and the lower off-diagonal panels show the 2D marginalized posteriors. This figure corresponds to the flavor-universal case, in which all three neutrino 
    flavors are self-interacting.}
    \label{fig:2D_posterior3I0f-cosmo}
\end{figure*}

\begin{figure*}[t]
    \centering
    \includegraphics[width=\textwidth]{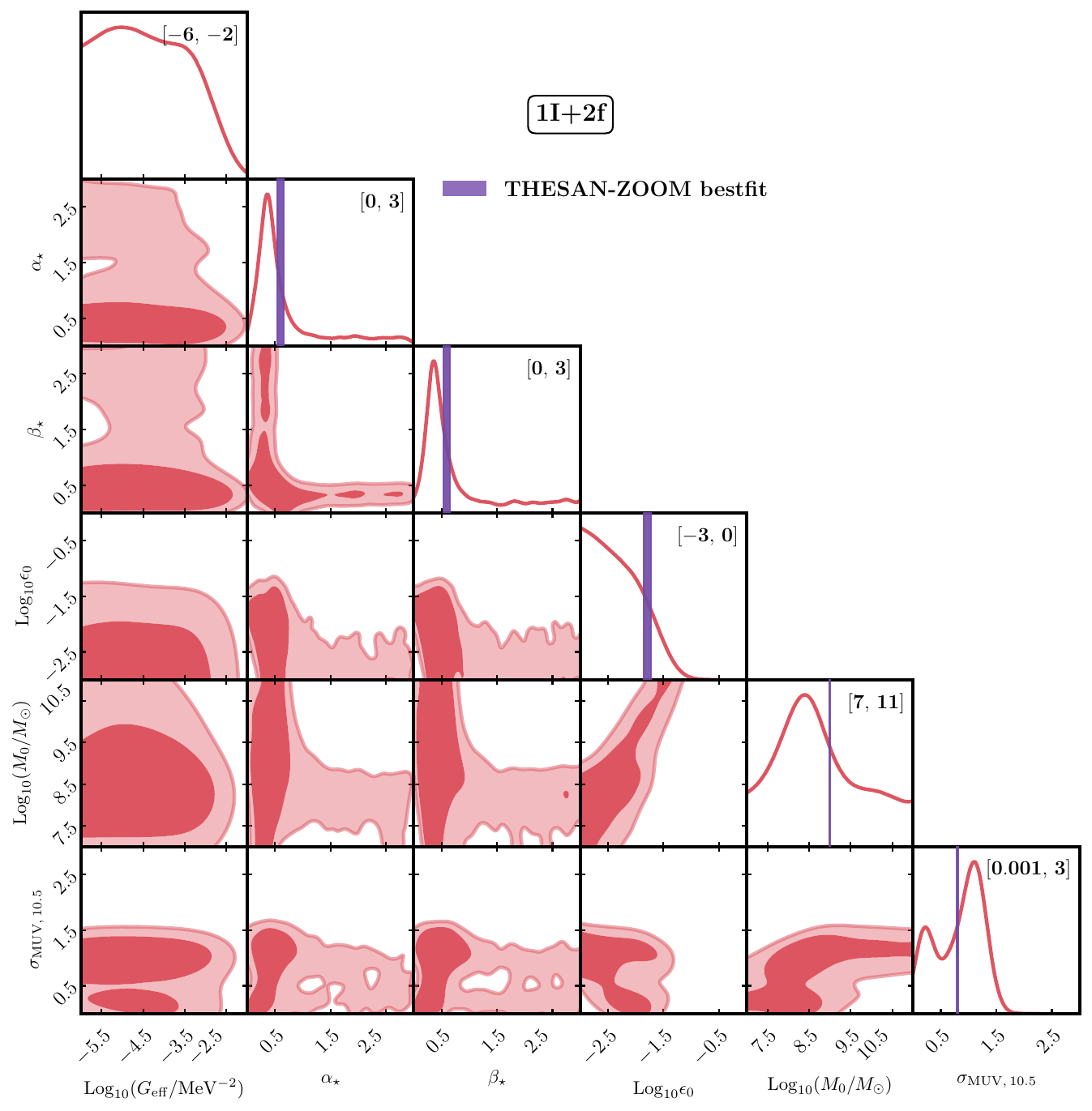}
    \caption{Same as Fig.~\ref{fig:2D_posterior3I0f}, but for flavor-specific case}
    \label{fig:2D_posterior1I2f}
\end{figure*}

The above discussion describes how $\nu\mathrm{SI}$ modifies the standard free-streaming picture, especially when the interaction strength is large. In this work, we focus on the MI regime, for which free streaming begins when most of the $\ell$ modes that we observe are still outside the horizon. However, this transition is not instantaneous, and the neutrino visibility function has a finite width~\cite{Cyr-Racine:2013jua, Kreisch:2019yzn}. Due to residual scattering during this transition, the full free-streaming-induced amplitude change and phase shift are not completely recovered, even in the MI regime~\cite{Kreisch:2019yzn}.

\begin{figure*}[t]
    \centering
    \includegraphics[width=\textwidth]{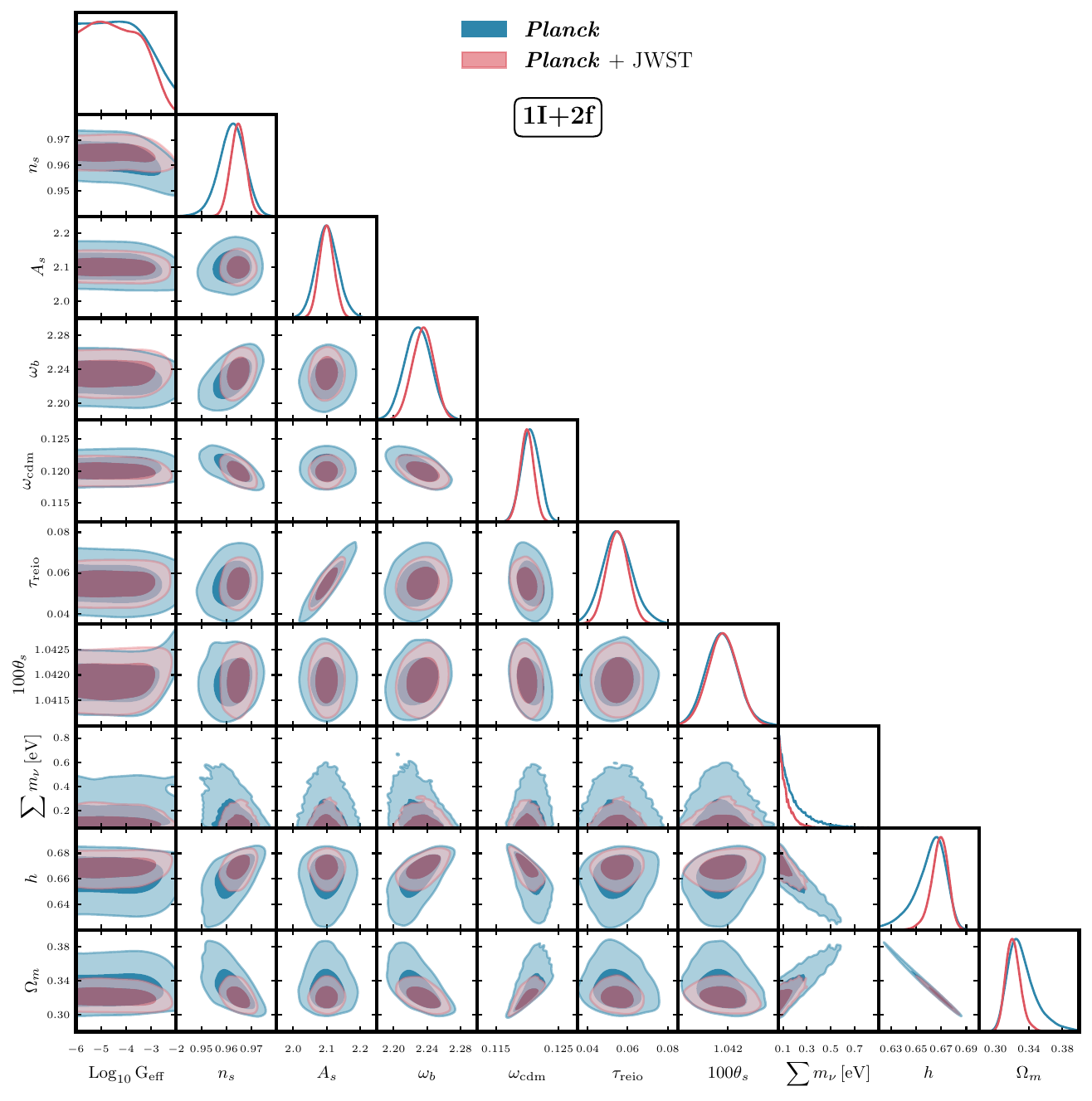}
    \caption{Same as Fig.~\ref{fig:2D_posterior3I0f-cosmo}, when only one-flavor is self-interacting.}
    \label{fig:2D_posterior1I2f-cosmo}
\end{figure*}

In Fig.~\ref{fig:planck_tt_ee_combined}, we show the CMB TT and EE power spectra against the \textit{Planck} 2018 data for the 95\% C.L. upper limit values of $\sum m_{\nu}$ and $G_{\mathrm{eff}}$ obtained from the \textit{Planck}-only analysis (blue shaded curves) and from the combined \textit{Planck}+JWST analysis (red shaded curves), for both the 1I + 2f and 3I + 0f cases. As expected, acoustic oscillations in $\nu$SI cosmologies show a relative phase shift towards larger $\ell$, accompanied by an enhancement in the CMB TT and EE spectra. The enhancement is more pronounced at larger multipoles, since these modes enter the horizon earlier and remain longer in the regime where residual effects of $\nu$SI persist.

Massive neutrinos affect the CMB power spectra through their impact on both the background evolution and the growth of structure~\cite{Herold:2024nvk, Lesgourgues:2012uu, Archidiacono:2017tlz, Loverde:2024nfi, Wong:2011ip, Archidiacono:2016lnv, Lesgourgues:2006nd}. The total non-relativistic matter density, $\omega_m=\Omega_m h^2$, receives a contribution from massive neutrinos and is therefore sensitive to the sum of the neutrino masses, $\sum m_\nu$. Changes in the matter density can affect the epoch of matter--radiation equality, $z_{\mathrm{eq}}$, as well as the angular diameter distance, $d_A$, to the last-scattering surface. They can also alter the epoch of matter--dark-energy equality and consequently modify the late-time integrated Sachs--Wolfe (ISW) contribution. In addition, massive neutrinos suppress structure formation at wavenumbers above a characteristic free-streaming scale. The characteristic wavenumber $k_{\rm nr}$ is associated with the epoch at which neutrinos become non-relativistic. The resulting suppression of structure reduces the gravitational lensing of CMB photons and thereby modifies the lensed CMB power spectra. Combining CMB observations with probes of structure across different length scales can therefore provide tighter constraints on $\sum m_\nu$. In this work, we also study the constraints on $\sum m_\nu$ obtained from the combined \textit{Planck}+JWST analysis. Despite the currently limited dataset and relatively large observational uncertainties, JWST can tighten the constraints on $\sum m_\nu$.

In Fig.~\ref{fig:planck_tt_ee_combined}, the changes in the TT and EE power spectra  due to increase in $\sum m_{\nu}$ mainly comes through changes in the background evolution, and the growth of perturbations, leading to a mild suppression and small shifts in the acoustic features.

Fig.~\ref{fig:neutrino-mass-posterior} shows 1D marginalized posteriors for $G_{\mathrm{eff}}$ (left panel) and $\sum m_{\nu}$ (right panel) obtained from the \textit{Planck} data (blue solid lines) and the \textit{Planck}+JWST data (red solid lines), for the 3I + 0f self-coupling case. The dashed vertical lines indicate the corresponding 95\% C.L. upper limits on the parameters, using the same color coding for the two datasets.\,\,We find that the posterior for $G_{\mathrm{eff}}$ from the \textit{Planck}+JWST data has a sharper drop compared to the posterior from the \textit{Planck} data alone. As discussed previously, this additional sensitivity arises because JWST disfavors any suppression of structure formation in the presence of $\nu$SI. Similarly, JWST also disfavours any suppression resulting from larger $\sum m_{\nu}$ at the relevant scales, providing a probe complementary to \textit{Planck}. This explains why R.H.S in Fig.~\ref{fig:neutrino-mass-posterior} and Fig.~\ref{fig:sum_mnu_1_flavor} adding JWST shrinks the posterior. Since both $G_{\mathrm{eff}}$ and $\sum m_{\nu}$ suppress the matter power spectrum on the relevant scales, increasing the lower prior bound on $\sum m_{\nu}$ leads to stronger constraints on $G_{\rm eff}$. We adopt a lower prior bound of $\sum m_{\nu}=0.06~\mathrm{eV}$, corresponding to the minimum value allowed by neutrino oscillation experiments.

Fig.~\ref{fig:2D_posterior3I0f} and Fig.~\ref{fig:2D_posterior1I2f} show the 1D and 2D marginalized posteriors for the astrophysical parameters. The posteriors indicate wide range of model parameters can be consistent with the UVLF measurements. There is no strong correlation between the $\mathrm{Log}_{10}(G_{\rm eff}/\rm MeV^{-2})$ and the other astrophysical parameters. Marginalizing over the astrophysical parameters ensures that our constraints are robust against variations in the filter and fitting functions. Figs. \ref{fig:2D_posterior3I0f-cosmo} and \ref{fig:2D_posterior1I2f-cosmo} shows the 1D and 2D marginalized posteriors on the cosmological parameters for \textit{Planck}-only (in blue) and \textit{Planck}+JWST (in red) analysis, for three flavor and one-flavor interacting cases, respectively.

\section{$\nu$SI Models}
We present our constraints on $G_{\mathrm{eff}}$ using the \textit{Planck} + JWST data, where $G_{\mathrm{eff}}$ denotes the effective coupling strength of the neutrino four-fermion  interaction~\cite{Blinov:2019gcj},
\begin{equation}
\mathcal{L} \supset  G_{\mathrm{eff}}^{\alpha\beta\gamma\delta}\bar{\nu}_{\alpha}\nu_{\beta}\bar{\nu}_{\gamma}\nu_{\delta},\label{eq:eff-interaction}
\end{equation}
where $\alpha$, $\beta$, $\gamma$, and $\delta$ are flavor indices. This effective description is valid when the momentum transfer is much smaller than the mediator mass in a UV-complete theory. For the cosmological epochs relevant to our analysis, eq.~\eqref{eq:eff-interaction} remains a good approximation for mediator masses of order $\sim 0.1\, \rm MeV$ or heavier. At higher energies, when the momentum transfer becomes comparable to or larger than the mediator mass, this contact-interaction approximation breaks down. A renormalizable theory that reproduces eq.~\eqref{eq:eff-interaction} as its effective description can be written in terms of a scalar mediator of mass $m_\phi$, coupled to neutrinos with strength $g_\phi$~\cite{Blinov:2019gcj}
\begin{figure}[t]
\begin{flushleft}
\includegraphics[width=1.04\columnwidth]{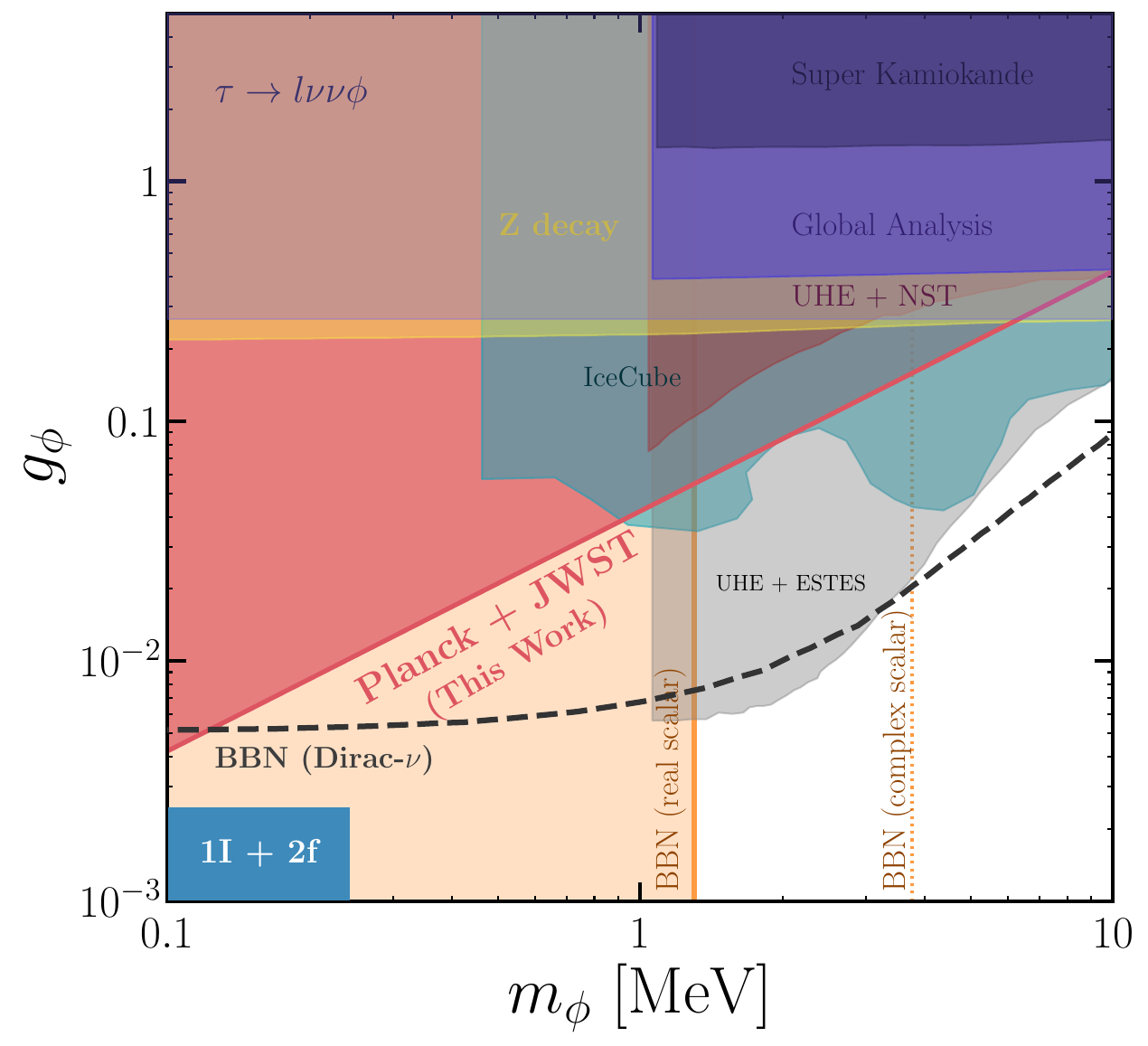}
\end{flushleft}
    \caption{Comparison of the 95\% C.L. constraints obtained from our analysis with the \textit{Planck}+JWST data (red shaded region) for flavor-specific coupling to $\nu_{\tau}$ with previous limits for a self-interaction model. For Majorana neutrinos with a real scalar mediator, our constraints are the strongest among the other cosmological probes beyond $m_{\phi} \sim 1\,\mathrm{MeV}$. Even when these conditions are relaxed, our limits remain competitive with other terrestrial limits, such as those from $\tau$ decay, and with cosmological limits from BBN.
}
    \label{fig:model-constraint}
\end{figure}

\begin{equation}
\mathcal{L} \supset -\frac{1}{2}m_{\phi}^{2}\phi^{2} + \frac{1}{2}\Big(g_{\phi}^{\alpha\beta}\nu_{\alpha}\nu_{\beta}\phi + \mathrm{h.c}\Big)\,,
\end{equation}
where we assume $\phi$ is a real scalar and $(\alpha,\beta) \in (e, \mu, \tau)$. Under the assumption $|q^{2}| \ll m_{\phi}^{2}$, $g_{\phi}$ and $m_{\phi}$ are related to the $G_{\mathrm{eff}}$ as
\begin{equation}
G_{\mathrm{eff}} = \frac{g_{\phi}^{2}}{m_{\phi}^{2}}\,,
\end{equation}
where $g^{\alpha\beta}_{\phi} \equiv g_{\phi}$. Fig.~\ref{fig:model-constraint} shows the parameter space excluded by our 95\% C.L. limit for 1I + 2f interactions from the \textit{Planck} + JWST analysis (red shaded region) onto the $g_{\phi}-m_{\phi}$ plane for the light scalar mediator model given in Ref.~\cite{Blinov:2019gcj}. We show our results against the other existing limits for flavor-specific couplings to $\nu_{\tau}$.\,\,The MI region is already strongly constrained for couplings involving $\nu_e$ and $\nu_\mu$ by laboratory probes. We therefore focus on the $\nu_\tau$ flavor-specific case, which is least constrained. Since cosmological observables are insensitive to the flavor of the interacting neutrino, the 1I + 2f cosmological constraint can be mapped onto any single-flavor coupling scenario. The other limits in the $g_{\phi}-m_{\phi}$ plane arise mainly from laboratory experiments, BBN, and the high energy astrophysical neutrinos~\cite{Blinov:2019gcj, Agostini:2015nwa, Blum:2018ljv, Brune:2018sab, Foroughi-Abari:2025mhj, He:2025bex, Leal:2025eou, Brdar:2020nbj}. For Majorana neutrinos and real scalar mediators, we place the \textit{strongest constraint} in a region around $m_{\phi}\sim\mathcal{O}(\mathrm{MeV})$ among the other cosmological observables. Our limits are also competitive with the other $\nu$SI constraints from the high-energy astrophysical neutrinos.

\begin{figure*}
	\begin{center}
		\includegraphics[width=0.98\columnwidth]{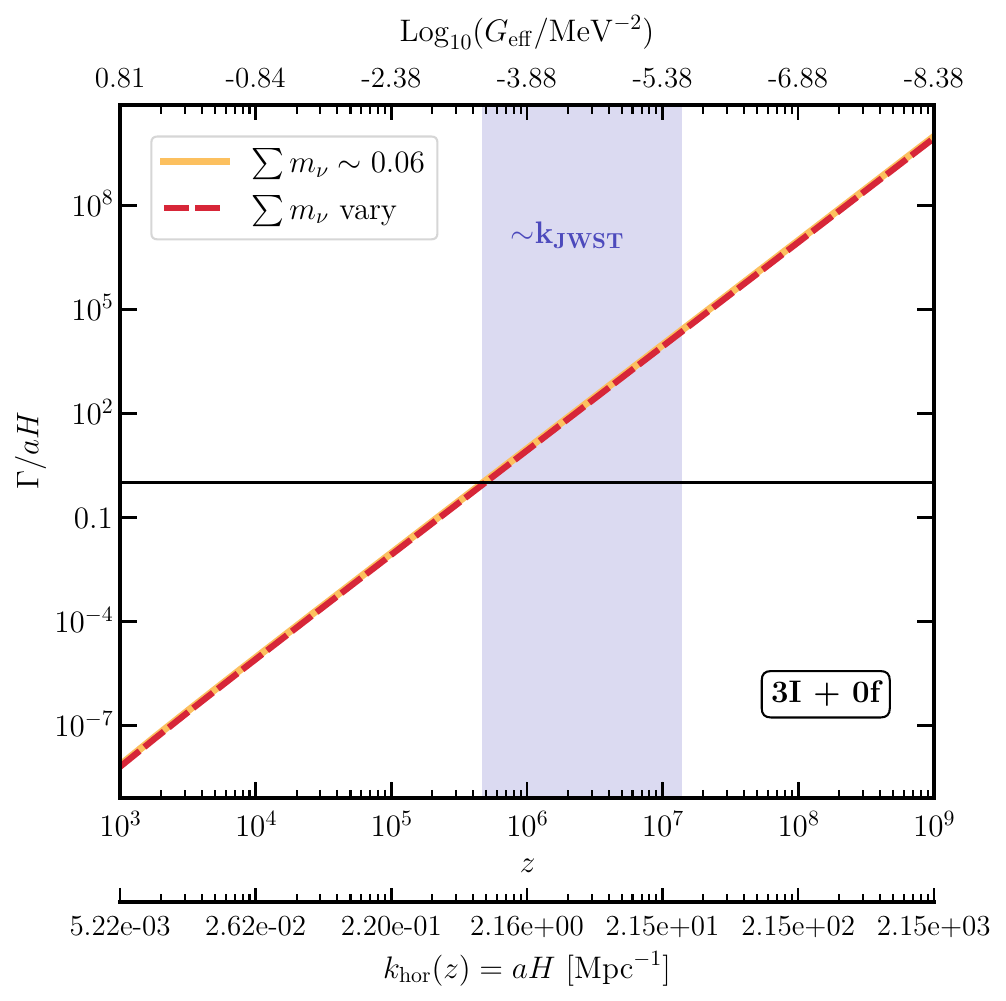}~~
		\includegraphics[width=0.98\columnwidth]{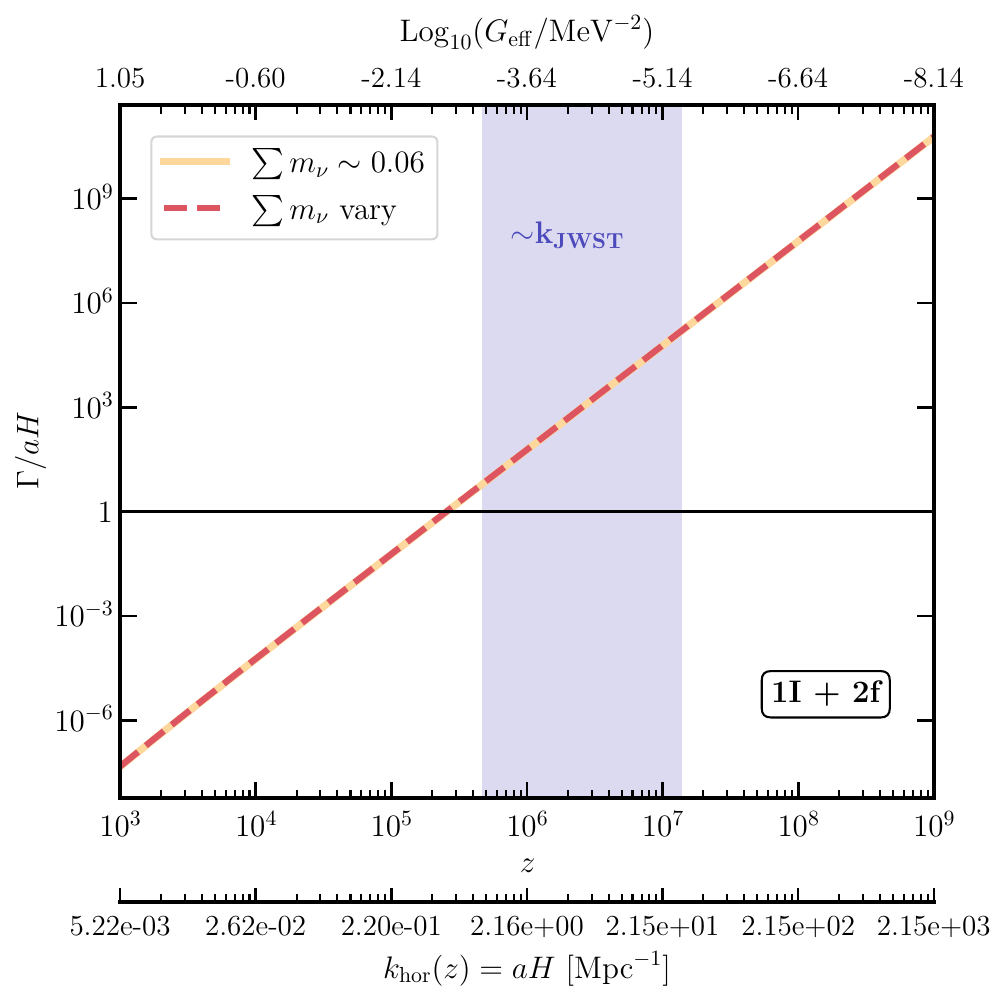}~~\\	
		\caption{
We show the ratio of the neutrino self-interaction rate to the conformal 
Hubble rate, $\Gamma/aH$, as a function of redshift. The horizontal black line 
marks $\Gamma=aH$, which approximately indicates the epoch of neutrino 
decoupling, or equivalently the onset of neutrino free streaming. The purple 
shaded band shows the approximate range of scales probed by JWST. The lower 
horizontal axis shows the comoving horizon wavenumber, 
$k_{\rm hor}(z)=aH$, corresponding to each redshift, while the upper horizontal 
axis shows the interaction strength for which neutrinos begin to free stream 
at the corresponding redshift. The solid orange and dashed red curves correspond 
to the 95\% C.L. upper limit of the interaction strength obtained for 
$\sum m_\nu \simeq 0.06\,{\rm eV}$ and varying $\sum m_\nu$, respectively. 
\textit{Left panel:} Results for the three-flavor interacting case, 
$3{\rm I}+0{\rm f}$. The dashed red curve corresponds to 
$\mathrm{Log}_{10}(G_{\rm eff}/{\rm MeV}^{-2})=-3.41$, obtained by varying 
$\sum m_\nu$, while the solid orange curve corresponds to 
$\mathrm{Log}_{10}(G_{\rm eff}/{\rm MeV}^{-2})=-3.39$, obtained from runs with fixed $\sum m_\nu \simeq 0.06\,{\rm eV}$. 
\textit{Right panel:} Results for the flavor-specific interacting case, 
$1{\rm I}+2{\rm f}$. The dashed red curve corresponds to 
$\mathrm{Log}_{10}(G_{\rm eff}/{\rm MeV}^{-2})=-2.75$, obtained by varying 
$\sum m_\nu$, while the solid orange curve corresponds to 
$\mathrm{Log}_{10}(G_{\rm eff}/{\rm MeV}^{-2})=-2.75$, obtained for 
$\sum m_\nu \simeq 0.06\,{\rm eV}$.
    }
		\label{fig:fs_redshift_both}
	\end{center}	
\end{figure*}

\section{Halo Accretion rates for modified scenarios}
Since the UVLF depends on the halo mass and the redshift, it also depends on the formation and subsequent evolution of DM halos. In $\Lambda$CDM cosmology, smaller halos form first and eventually merge or accrete matter to form larger halos.  The HMF depends on the matter power spectrum, $P(k)$, and therefore the parametric accretion-rate model, calibrated using CDM-only simulations, may not provide an exact description of halo accretion in cosmologies with a modified $P(k)$~\cite{2013MNRAS.428.1774B, Lovell:2013ola}.

The impact of changes in $P(k)$ on the halo accretion rate has been studied in the context of the WDM models~\cite{Lovell:2013ola}. Both semi-analytical analyses and simulations have been performed to study the effect of WDM cosmology on the accretion rate. For models that introduce a cutoff scale in $P(k)$, thereby suppressing the abundance of low-mass halos, the resulting change in halo abundance is expected to affect merger-driven accretion. This effect becomes important below a certain halo mass, where smaller halos that would otherwise contribute to growth through mergers become less abundant~\cite{2013MNRAS.428.1774B}.

As emphasized in~\cite{2013MNRAS.428.1774B}, even for WDM scenarios, where both the halo mass function and the progenitor mass function change substantially with respect to CDM, the mean halo accretion rate does not show any difference. $\nu \mathrm{SI}$ does not introduce a cutoff scale in $P(k)$. Instead, they induce an enhancement at large scales followed by a suppression at smaller scales, at most at the level of 15\%. We therefore expect the corresponding deviations in the mean halo accretion rate to be even smaller than those found in WDM scenarios. Hence, adopting the same accretion physics as calibrated from the \textsc{GUREFT} simulations gives us a fairly accurate scenario in this context.

\section{What lies in the future?}

JWST has been operating for about four years, observing structure formation at very high redshifts. At these high redshifts, there are not many DM halos, and even fewer bright structures. This is because the brightness comes from the stars inside galaxies, which are more likely to form if a halo is massive enough to accrete baryons and form stars. This is what caused a stir in the astrophysics community when JWST started to observe more galaxies at the bright end of the UV luminosity functions. Nevertheless, the number of galaxies observed is still very small, resulting in large error bars~\cite{2025ApJ...980..138H}. Despite these limitations, we would like to understand the potential of JWST and other high-redshift observables to probe new physics. It is hard to predict how much the error bars will shrink in future JWST UVLF measurements without performing a dedicated computation, since they depend on both the number of galaxies at these high redshifts and the capabilities of the telescope.

Instead of trying to do a forecast, we ask how far we can push the limits if we extend our argument regarding suppression due to $\nu$SI to high-$k$ observations. Here, we include JWST photometric UVLF measurements up to redshift $z\sim25$. Structure formation is hierarchical in nature, and the relevant halo masses at these redshifts correspond to small comoving scales, roughly extending to $k\sim30\,\mathrm{Mpc}^{-1}$~\cite{Nadler:2025fcv}. Each $k$ mode can be associated with the redshift at which it enters the horizon. Similarly, each redshift can be associated with an interaction strength for which the onset of free streaming occurs around that redshift. If JWST observes galaxies that can be associated with structures at larger $k$, then it would disfavour a neutrino-interaction model that predicts suppression at those $k$ values.

In Fig.~\ref{fig:fs_redshift_both}, we plot the ratio of the interaction rate $\Gamma$ to the conformal Hubble rate $aH$ as a function of redshift for both the flavor-universal (left panel) and flavor-specific (right panel) interactions. The lower $x$-axis shows the horizon-entry scale, $k_{\rm hor}(z)=aH$, associated with each redshift. The upper $x$-axis shows the corresponding interaction strength, $\mathrm{Log}_{10}(G_{\rm eff}/\mathrm{MeV}^{-2})$, for which neutrino decoupling occurs around the redshifts shown in the lower panel, marking the onset of neutrino free streaming. In both panels, $\Gamma/aH$ is evaluated for interaction strengths fixed to the 95\% C.L. upper limits obtained from the \textit{Planck}+JWST likelihoods for two different cases. The red dashed curve corresponds to the case where $\sum m_{\nu}$ is varied as an MCMC parameter, while the orange curve represents a case where $\sum m_{\nu}$ is fixed to $0.06\,{\rm eV}$. As we decrease $G_{\mathrm{eff}}$, the straight lines shift towards the right, and the onset of free streaming moves to higher redshifts, indicating earlier free streaming, with structure formation being affected for $k>k_{\rm fs}$. If the UVLF measurements extend to higher redshifts, indicating the presence of structures at even earlier epochs, then the interaction strengths that leads to a suppression around \(k\sim 30\,{\rm Mpc}^{-1}\) can be excluded. This allows us to probe values of $\mathrm{Log}_{10}(G_{\mathrm{eff}}/\mathrm{MeV}^{-2})$ as low as $\sim -5$. However, as we see from our $T^{2}(k)$ plots, the effect of neutrino self-interactions does not lead only to suppression, but can also produce an enhancement. Determining which feature plays the dominant role requires a more quantitative analysis and is beyond the scope of this work.

\end{document}